\documentclass[12pt,fleqn]{article}
\usepackage{amsmath,amssymb,bbm,amsthm,tikz,indentfirst,graphicx,mathrsfs,enumerate,epstopdf,caption,subcaption,float,authblk}
\usepackage[square,comma,numbers,sort&compress]{natbib}
\usepackage[colorlinks]{hyperref}
 \theoremstyle{plain}
\newtheorem{theorem}{Theorem}[section]

\newtheorem{lemma}[theorem]{Lemma}

 \theoremstyle{remark}
\newtheorem{remark}[theorem]{Remark}
 \theoremstyle{definition}
 \newtheorem{definition}[theorem]{Definition}
\newtheorem{problem}[theorem]{RH Problem}
\usepackage[left=2.5cm,right=2.5cm,bottom=3cm,top=3cm]{geometry}

\newcommand{\res}{\operatorname{Res}\ }
\newcommand{\ii}{\mathrm{i}}
\newcommand{\dd}{\mathrm{d}}
\newcommand{\e}{\mathrm{e}}
\newcommand{\T}{\mathrm{T}}
\newcommand{\ut}{\underline{\mathbf{t}}}

\numberwithin{equation}{section}
\allowdisplaybreaks[2]
\begin{document}
\title{ Generalized discrete integrable operator and   integrable hierarchy}

\renewcommand*{\Affilfont}{\small\it}
\renewcommand{\Authands}{, }
\author[1]{Huan Liu\thanks{ E-mail: liuhuan@zzu.edu.cn}\\School of Mathematics and Statistics, Zhengzhou University, Zhengzhou, Henan 450001, China}
\date{}
  \maketitle
\begin{abstract}

We introduce and systematically develop two classes of discrete integrable operators: those with $2\times 2$ matrix kernels and those possessing general differential kernels, thereby generalizing the discrete analogue previously studied. A central finding is their inherent connection to higher-order pole solutions of integrable hierarchies, contrasting sharply with standard operators linked to simple poles. This work not only provides explicit resolvent formulas for matrix kernels and differential operator analogues but also offers discrete integrable structures that encode higher-order behaviour.
\end{abstract}
\section{Introduction}
  In  mathematical physics, {\it integrability} stands as a pivotal concept for characterizing the solvability  of complex systems. Numerous challenging tasks, spanning from calculating correlation functions in quantum integrable models and analyzing spectral distributions in random matrix theory, to describing probabilistic behaviors in discrete combinatorics and obtaining exact solutions for nonlinear partial differential equations (PDEs), have historically depended on uncovering operator classes possessing special algebraic and analytical properties.  Among these,  the IIKS integrable operators, systematically proposed by Its, Izergin, Korepin, and Slavnov \cite{IIKS1990}, stand out. They offer a unified analytical framework across diverse fields through their distinctive kernel structure and deep connections to Riemann--Hilbert (RH) problems. This framework has proven instrumental in expanding the applicability of integrable systems theory over the past three decades \cite{DP1993}.

 The foundational work on IIKS operators primarily addressed integral operators defined on continuous contours. A landmark achievement was recognizing that the resolvent of such operators could be determined by solving an associated RH problem \cite{IIKS1990,IIKS1993,HI2002}. This crucial insight transformed  the inverse problem of infinite-dimensional operators into a finite-dimensional complex analysis problem, providing exact tools for calculating correlation functions in quantum integrable models. However, the strict analyticity inherent in traditional RH problems proved insufficient for tackling certain nonlinear PDEs with non-analytic initial data. Addressing this,  Bertola et al. further extended the IIKS framework to $\bar\partial$-problems \cite{BGO2024}, defining integrable operators on bounded complex domains. These operators retain the IIKS kernel structure but replace the analyticity condition with a constraint given by a $\bar\partial$-equation. 
 
Simultaneously, many practical problems in areas like combinatorics and discrete lattice models demand operators defined on discrete spaces, motivating the development of discrete integrable operators.  Borodin \cite{Borodin2000} pioneered this direction by extending the IIKS framework to discrete settings, introducing the discrete RH problem. Unlike the continuous case, this requires matrix-valued functions with simple poles, imposing additional constraints through residue conditions.

  Discrete integrable operators play a pivotal role in analyzing correlation functions of determinantal point processes arising in representation theory, random matrix theory and integrable system. Notable examples include the discrete Bessel kernel, capturing longest increasing subsequences in permutations \cite{BOO2000}, and the confluent hypergeometric kernel, governing eigenvalue statistics in random matrices \cite{BO2000,BD2002,Borodin2003}. Furthermore, these operators enable exact computations of spectral measures for infinite symmetric groups via RH problems \cite{BOO2000,BO2000,Borodin2003,BD2002}.  As highlighted in Cafasso and Ruzza  \cite{CR2023}, they also characterize the finite-temperature deformation of the discrete Bessel point process on $\mathbb{Z}+\frac12$, linking to 2D Toda reductions and discrete Painlev\'{e} equation, and continuum limits connecting to Airy kernels and Korteweg--de Vries (KdV) equation.

In addition, the significance of discrete integrable operators extends to the study of universal fluctuations, particularly within the Kardar--Parisi--Zhang (KPZ) universality class.  They provide a critical analytical bridge between integrable systems and KPZ fluctuations. For instance, Baik et al. \cite{BLS2022,BPS2023} apply these operators to the periodic KPZ fixed point, acting on discrete Bethe roots and representing multi-point distributions via Fredholm determinants built from kernels with cubic exponential factors, leading to coupled matrix nonlinear Schr\"odinger (NLS) and modified KdV (mKdV) equations. Quastel and Remenik \cite{QR2022} show that Fredholm determinants involving these operators govern the matrix Kadomtsev--Petviashvili (KP) equation for KPZ fixed-point multi-point distributions, with Tracy--Widom distributions appearing as  self-similar solutions.
Beyond the role of discrete integrable operators, integrable structures more broadly imply profound intrinsic connections exist between integrable equations and the universal fluctuations of the KPZ universality class \cite{CCR2022,MQR2025,Krajenbrink2020,KL2021,MMS2022}.
   
These diverse findings collectively illustrate the significance of both continuous and discrete integrable operators. Building upon this foundation, this study addresses three key questions stemming from natural extensions and current research directions:
   
   First,  $2\times 2$ matrix kernels are pivotal in  random matrix theory for analyzing eigenvalue correlations in orthogonal, symplectic ensembles and  infinite Gelfand pair \cite{BS2009,GP2002,BBCS2018,Kargin2014,Petrov2011,FM2012,Strahov2010a,Strahov2010b,Forrester2010,FL2020}. These kernels naturally emerge in the Pfaffian representations of $m$-point correlation functions  but are traditionally expressed via skew orthogonal polynomials, which often lack explicit forms and require intensive evaluations of Christoffel--Darboux sums.  To address this, researchers have used classical orthogonal polynomials with the same weight function for their established integrability properties \cite{Widom1999} and deep connections to RH problems \cite{Little2024}.  Solving the RH factorization for these kernels constructs explicit resolvent operators.  Extending the framework of integrable operators to  $2\times 2$  matrix kernels raises a critical question:  does the integrable structure, including both the RH factorization mechanism and resolvent formulae, persist under matrix generalization?
 
 Second,  inspired by the factorization of Fredholm determinants for shifted integral kernels by Its and Kozlowski \cite{IK2014,IK2016}, which effectively replaces an operator-valued RH problem with scalar matrix factorization using a discrete difference analogy, we ask whether analogous integrable operators exist with differential matrix kernels.

 Third, in the classical setting, the original discrete integrable operator encodes the RH problem for pure soliton solutions, whose transmission coefficients have only simple poles in the spectral plane. Recent extensions handle higher-order pole solutions \cite{LSG20241,LSG20242,Li2025},  whose transmission coefficients exhibit higher-order singularities. A subtle and open question is whether a discrete integrable operator can be constructed to match these higher-order solutions.

To tackle these questions, this paper aims to expand the class of discrete integrable operators by incorporating integrable matrix-type and differential-type operators and establishing their connection to higher-order pole solutions of integrable hierarchies. We organize the paper as follows: Section 2  introduces the necessary functional analytic framework and reviews continuous and discrete integrable operators.  Section 3 introduces $2\times2$ matrix-kernel discrete integrable operators. Section 4 derives this class of integrable matrix kernels associated with orthogonal polynomials. Section 5 presents general discrete integrable operators with differential kernels. Section 6 demonstrates the construction of higher-order pole solutions  for integrable hierarchies using these operators.

\section{A  brief view of integrable operators}
Introduce a subset $\Omega \subset \mathbb{C}$ and an associated measure $\mu$ to construct the space $L^2(\Omega, \mu)$. When $\Omega$ is a discrete set without accumulations points, the measure $\mu$ is defined as the counting measure, and $L^2(\Omega, \mu)$ becomes $\ell^2(\Omega)$, the space of square-summable sequences. Conversely, if $\Omega$ is a finite union of disjoint simple contours, the measure is set as $d\mu = ds$, corresponding to integration along the contours, and $L^2(\Omega, \mu)$ becomes $L^2(\Omega, ds)$, the space of square-integrable functions. This framework provides a unified approach to analyzing integrable operators in both discrete and continuous settings, ensuring consistency and mathematical rigor.

\begin{definition}
  Define $\mathsf{K}:L^2(\Omega, \mu)\rightarrow L^2(\Omega, \mu)$   by
\begin{equation}
 \mathsf{K}[h](\xi)=\int_\Omega K (\xi,\eta)h(\eta)\dd\mu(\eta),\quad K(\xi,\eta)=\frac{\mathbf{f}^\T(\xi)\mathbf{g}(\eta)}{\xi-\eta},
\end{equation}
 where $\mathbf{f},\mathbf{g}:\mathbb{C}\rightarrow\mathbb{C}^{n+1}$ are smooth column vector-valued functions. 
\end{definition}
It should be noted that the superscript T denotes the transpose of a matrix, while $-$T represents the inverse transpose of a matrix.
 To ensure the kernel  $K$ is nonsingular, we impose the condition
$\mathbf{f}^\T(\xi)\mathbf{g}(\xi)=0$ for $\xi\in\Omega$. This allows the diagonal values of $K$  to be defined via the limit:
\begin{equation}
  K(\xi,\xi)=\partial_\xi[\mathbf{f}^\T(\xi)]\mathbf{g}(\xi)=-\mathbf{f}^\T(\xi)\partial_\xi[\mathbf{g}(\xi)],\quad   \xi\in\Omega.
\end{equation}
 In the discrete case, we additionally require   $\partial_\xi[\mathbf{f}^\T(\xi)]\mathbf{g}(\xi)=-\mathbf{f}^\T(\xi)\partial_\xi[\mathbf{g}(\xi)]=0$ to force the diagonal kernel to vanish.
The operator $\mathsf{K}$ is understood as acting on the right, while  its transpose $\mathsf{K}^\T$
acts on the left:
\begin{equation}
 \mathsf{K}^\T[h](\eta)=\int_\Omega h(\xi)K(\xi,\eta)\dd\mu(\xi),\quad \eta\in\Omega.
\end{equation}

A first important observation to note   is that the resolvent operator
$\mathsf{R}=(\mathbbm{1}-\mathsf{K})^{-1}-\mathbbm{1}$
is also in the same class.
\begin{theorem}(Its–Izergin–Korepin–Slavnov  \cite{IIKS1990,IIKS1993} $\&$ Borodin \cite{Borodin2000})
  If  $\mathbbm{1}-\mathsf{K}$ is invertible,  the resolvent operator $
  \mathsf{R}:L^2(\Omega, \mu)\rightarrow L^2(\Omega, \mu)$
 has the  kernel
\begin{equation}\label{def:R}
   R(\xi,\eta)=\frac{\mathbf{F}^\T(\xi)\mathbf{G}(\eta)}{\xi-\eta},
\end{equation}
where
\begin{subequations}
\begin{align}
 & \mathbf{F}^\T=(\mathbbm{1}-\mathsf{K})^{-1}\mathbf{f}^\T,\label{reFf}\\
  &\mathbf{G}=(\mathbbm{1}-\mathsf{K}^\T)^{-1}\mathbf{g}.\label{reGg}
\end{align}
\end{subequations}
\end{theorem}

The second principal observation of \cite{IIKS1990,IIKS1993,Borodin2000} is that the resolvent kernel $R$ can be recovered from the solution of the following RH problem.
\begin{problem}\label{RHIIKS}
Find a matrix-vauled function $\mathsf{M}(\xi)\in \mathbb{C}^{(n+1)\times(n+1)}$ such that
  \begin{enumerate}[(i)]
    \item $\mathsf{M}(\xi)$ is analytic on $ \mathbb{C}\backslash\Omega$.
    \item As $\xi\rightarrow\infty$, $\mathsf{M}(\xi)=\mathbf{I}+O(\xi^{-1})$.
    \item The jump/pole behaviour of  $\mathsf{M}(\xi)$  depends on whether $\Omega$ is a union of contours or a discrete set.
    \begin{itemize}
      \item continuous case:    for any point $\xi\in\Omega$, the boundary values of  $\mathsf{M}(\xi)$ from opposite sides of the oriented contour are related by
\begin{equation}
   \mathsf{M}_+(\xi) = \mathsf{M}_-(\xi)[\mathbf{I} - 2\pi \ii \mathbf{f}(\xi) \mathbf{g}^\T(\xi)],
\end{equation}
      \item discrete case:  every point $\kappa\in\Omega$ is the simple pole of $\mathsf{M}(\xi)$, the residues at these poles are determined by
    \begin{equation}
      \underset{\xi=\kappa}{\res}\mathsf{M}(\xi)=\lim_{\xi\rightarrow\kappa}\mathsf{M}(\xi)\mathbf{f}(\kappa)\mathbf{g}^\T(\kappa).
            \end{equation}
    \end{itemize}
  \end{enumerate}
\end{problem}
Because $\left[\mathbf{f}(\xi) \mathbf{g}^\T(\xi)\right]^2=\mathbf{0}$, one has $\det[\mathsf{M}(\xi)]\equiv1$. Hence the solution to  RH problem \ref{RHIIKS}, if it exists, is unique and invertible.
\begin{theorem}(\cite{IIKS1990,IIKS1993,Borodin2000})
If  $\mathbbm{1}-\mathsf{K}$ is invertible.  RH problem \ref{RHIIKS} admits the unique solution
   \begin{equation}
       \mathsf{M}(\xi)=\mathbf{I}+\int_{\Omega}\frac{\mathbf{F}(\eta)\mathbf{g}^\T(\eta)}{\xi-\eta}\dd\mu(\eta),\quad\xi\notin\Omega,
   \end{equation}
  with $\mathbf{F}$ given by \eqref{reFf}. Conversely, if RH problem \ref{RHIIKS} is uniquely solved by $ \mathsf{M}(\xi)$, the resolvent kernel is
\begin{equation}
   R(\xi,\eta)=\frac{\mathbf{F}^\T(\xi)\mathbf{G}(\eta)}{\xi-\eta},  \quad \mathbf{F}^\T(\xi)\mathbf{G}(\xi)=0,
\end{equation}
where
\begin{itemize}
\item continuous case:
\begin{equation}
  \begin{aligned}
 &\mathbf{F}(\xi)=\mathsf{M}(\xi)\mathbf{f}(\xi),\quad
  \mathbf{G}(\eta)=\mathsf{M}^{-\T}(\eta)\mathbf{g}(\eta),\\
 & R(\xi,\xi)=\partial_\xi[\mathbf{F}^\T(\xi)]\mathbf{G}(\xi)=-\mathbf{F}^\T(\xi)\partial_\xi[\mathbf{G}(\xi)],
 \end{aligned}
\end{equation}
\item discrete case:
\begin{equation}
\begin{aligned}
  &\mathbf{F}(\xi)=\lim_{\zeta\rightarrow\xi}\mathsf{M}(\zeta)\mathbf{f}(\xi),\quad
  \mathbf{G}(\eta)=\lim_{\zeta\rightarrow\eta}\mathsf{M}^{-\T}(\zeta)\mathbf{g}(\eta),\\
  &  R(\xi,\xi)=\left[\lim_{\zeta\rightarrow\xi}\partial_\zeta[\mathsf{M}(\zeta)]\mathbf{f}(\xi)\right]^\T\mathbf{G}(\xi).
\end{aligned}
 \end{equation}
\end{itemize}
\end{theorem}

\section{Integrable matrix-type  kernel }\label{sec:sdi}

 In this section, we introduce a class of discrete integrable operators whose kernels are  $2\times 2$ matrices. Inspired by  the seminal works of Its--Izergin--Korepin--Slavnov \cite{IIKS1990,IIKS1993} and Borodin \cite{Borodin2000}, we define the matrix-type operator $\mathcal{K}$ to be \emph{discrete integrable} when  it satisfies the following two characteristic properties:
\begin{enumerate}[(a)]
  \item The resolvent of $\mathcal{K}$ preserves the same structural form as $\mathcal{K}$ itself (Theorem \ref{TH:th1}).
  \item The operator $\mathbbm{1}-\mathcal{K}$ is invertible if and only if the associated RH problem has a unique solution (Theorem \ref{TH:RHiff}).
\end{enumerate}

\begin{definition}
  Introduce the operator
 \begin{equation}
   \mathcal{K}:l^2(\Omega)\otimes  \mathbb{C}^{2\times m} \rightarrow l^2(\Omega) \otimes \mathbb{C}^{2\times m}.
 \end{equation}
Twisted acting on a  $2\times m$ matrix function
 \begin{equation}
    \begin{pmatrix}
               \mathbf{h}_0^\T(\xi)\\
               \mathbf{h}_1^\T(\xi)
             \end{pmatrix},\quad \mathbf{h}_0, \mathbf{h}_1:\Omega\rightarrow\mathbb{C}^m,
 \end{equation} $\mathcal{K}$ is defined by the discrete integral transform
\begin{equation}
  \mathcal{K}\begin{pmatrix}
               \mathbf{h}_0^\T \\
               \mathbf{h}_1^\T
             \end{pmatrix}(\xi)=\sum_{\eta\in\Omega}\mathbf{K}(\xi,\eta)\begin{pmatrix}
               \mathbf{h}_1^\T(\eta) \\
               \mathbf{h}_0^\T(\eta)
             \end{pmatrix},\quad \xi\in\Omega,
\end{equation}
 whose kernel  $\mathbf{K}(\xi,\eta)\in \mathbb{C}^{2\times 2}$ satisfies $\mathbf{K}(\xi,\xi)=\mathbf{0}$ and, for $\xi\neq \eta$,
\begin{equation}\label{eq:mkk}
  \mathbf{K}(\xi,\eta)=
    \begin{pmatrix}
                            \frac{\mathbf{f}_0^\T(\xi)\mathbf{g}_0(\eta)}{\xi-\eta} & \frac{\mathbf{f}_0^\T(\xi)\mathbf{g}_1(\eta)}{\xi-\eta}+\frac{\mathbf{f}_0^\T(\xi)\mathbf{g}_0(\eta)}{(\xi-\eta)^2} \\
                            \frac{\mathbf{f}_1^\T(\xi)\mathbf{g}_0(\eta)}{\xi-\eta}-\frac{\mathbf{f}_0^\T(\xi)\mathbf{g}_0(\eta)}{(\xi-\eta)^2} &  \frac{\mathbf{f}_1^\T(\xi)\mathbf{g}_1(\eta)}{\xi-\eta}+ \frac{\mathbf{f}_1^\T(\xi)\mathbf{g}_0(\eta)-\mathbf{f}_0^\T(\xi)\mathbf{g}_1(\eta)}{(\xi-\eta)^2}-\frac{2\mathbf{f}_0^\T(\xi)\mathbf{g}_0(\eta)}{(\xi-\eta)^3}
                          \end{pmatrix}.
\end{equation}
Here $\mathbf{f}_j, \mathbf{g}_j:\mathbb{C}\rightarrow\mathbb{C}^{n+1} (j=0,1)$ are smooth column vector-valued functions.
\end{definition}

  To ensure that the kernel  $\mathbf{K}(\xi,\eta)$ remains  nonsingular, we impose the orthogonality condition
\begin{equation}\label{eq:noncon}
  \mathbf{f}_j^\T(\xi)\mathbf{g}_k(\xi)=\mathbf{g}_k^\T(\xi)\mathbf{f}_j(\xi)=0, \quad \xi\in\Omega, \quad j,k=0,1.
\end{equation}
 The transpose operator
\begin{equation}
  \mathcal{K}^\T:l^2(\Omega)\otimes \mathbb{C}^{m\times 2}  \rightarrow l^2(\Omega) \otimes \mathbb{C}^{m\times 2}
\end{equation} 
 twistedly acts on  the left for $(\mathbf{h}_0,\mathbf{h}_1)$   as follows:
\begin{equation}
[\mathcal{K}^\T(
               \mathbf{h}_0,\mathbf{h}_1)](\eta)=\sum_{\xi\in\Omega}(
               \mathbf{h}_1(\xi),\mathbf{h}_0(\xi)
             )\mathbf{K}(\xi,\eta),\quad \eta\in\Omega.
\end{equation}

 \begin{theorem}\label{TH:th1}
 Assume the operator  $\mathbbm{1}-\mathcal{K}$ is invertible. Then the resolvent operator $
  \mathcal{R}=(\mathbbm{1}-\mathcal{K})^{-1}-\mathbbm{1}$ is determined by the kernel
  \begin{equation}\label{def:R1}
     \mathbf{R}(\xi,\eta)=    \begin{pmatrix}
                            \frac{\mathbf{F}_0^\T(\xi)\mathbf{G}_0(\eta)}{\xi-\eta} & \frac{\mathbf{F}_0^\T(\xi)\mathbf{G}_1(\eta)}{\xi-\eta}+\frac{\mathbf{F}_0^\T(\xi)\mathbf{G}_0(\eta)}{(\xi-\eta)^2} \\
                            \frac{\mathbf{F}_1^\T(\xi)\mathbf{G}_0(\eta)}{\xi-\eta}-\frac{\mathbf{F}_0^\T(\xi)\mathbf{G}_0(\eta)}{(\xi-\eta)^2} &  \frac{\mathbf{F}_1^\T(\xi)\mathbf{G}_1(\eta)}{\xi-\eta}+ \frac{\mathbf{F}_1^\T(\xi)\mathbf{G}_0(\eta)-\mathbf{F}_0^\T(\xi)\mathbf{G}_1(\eta)}{(\xi-\eta)^2}-\frac{2\mathbf{F}_0^\T(\xi)\mathbf{G}_0(\eta)}{(\xi-\eta)^3}
                          \end{pmatrix},
  \end{equation}
  where
  \begin{equation}\label{FfGg2}
    \begin{pmatrix}
      \mathbf{F}_0^\T \\
      \mathbf{F}_1^\T
    \end{pmatrix}=(\mathbbm{1}-\mathcal{K})^{-1}\begin{pmatrix}
      \mathbf{f}_0^\T \\
      \mathbf{f}_1^\T
    \end{pmatrix},\quad (
      \mathbf{G}_0,\mathbf{G}_1
   )=(\mathbbm{1}-\mathcal{K}^\T)^{-1}(
      \mathbf{g}_0,\mathbf{g}_1).
  \end{equation}
  \begin{proof}
Let $ \mathcal{R}$ be defined by \eqref{def:R1}. We shall verify the resolvent identity
    \begin{equation}\label{eq:KsR}
      \mathbf{K}\circ\mathbf{R}=\mathbf{R}-\mathbf{K},
    \end{equation}
where the composition reads
  \begin{equation}
    (\mathbf{K}\circ\mathbf{R})(\xi,\eta)=\sum_{\zeta\in\Omega}\mathbf{K}(\xi,\zeta)\begin{pmatrix}
                                                                                                                   0 & 1 \\
                                                                                                                   1 & 0
                                                                                                                 \end{pmatrix}\mathbf{R}(\zeta,\eta).
  \end{equation}
To illustrate the argument we check the $(1,2)$-entry of the formula \eqref{eq:KsR}, the remaining entries are handled analogously.  Expanding the sum gives
  \begin{equation}\label{eq:KsR12}
  \begin{aligned}
        &(\mathbf{K}\circ\mathbf{R})_{12}(\xi,\eta)\\
        =&\sum_{\zeta\in\Omega}\left\{\frac{\mathbf{f}_0^\T(\xi)\mathbf{g}_0(\zeta)}{\xi-\zeta}\left[ \frac{\mathbf{F}_1^\T(\zeta)\mathbf{G}_1(\eta)}{\zeta-\eta}+ \frac{\mathbf{F}_1^\T(\zeta)\mathbf{G}_0(\eta)-\mathbf{F}_0^\T(\zeta)\mathbf{G}_1(\eta)}{(\zeta-\eta)^2}-\frac{2\mathbf{F}_0^\T(\zeta)\mathbf{G}_0(\eta)}{(\zeta-\eta)^3}\right]\right.\\
        &+\left.\left[\frac{\mathbf{f}_0^\T(\xi)\mathbf{g}_1(\zeta)}{\xi-\zeta}+\frac{\mathbf{f}_0^\T(\xi)\mathbf{g}_0(\zeta)}{(\xi-\zeta)^2}\right]\left[\frac{\mathbf{F}_0^\T(\zeta)\mathbf{G}_1(\eta)}{\zeta-\eta}+\frac{\mathbf{F}_0^\T(\zeta)\mathbf{G}_0(\eta)}{(\zeta-\eta)^2}\right]\right\}.
  \end{aligned}
  \end{equation}
Inserting  the fraction identities
  \begin{equation}\label{eq:lmpow}
    \begin{aligned}
      &\frac{1}{\xi-\zeta}\frac{1}{\zeta-\eta}=\frac{1}{\xi-\eta}\left(\frac{1}{\xi-\zeta}+\frac{1}{\zeta-\eta}\right),\\
      &\frac{1}{\xi-\zeta}\frac{1}{(\zeta-\eta)^2}=\frac{1}{(\xi-\eta)^2}\left(\frac{1}{\xi-\zeta}+\frac{1}{\zeta-\eta}\right)+\frac{1}{\xi-\eta}\frac{1}{(\zeta-\eta)^2},\\
&\frac{1}{(\xi-\zeta)^2}\frac{1}{\zeta-\eta}=\frac{1}{(\xi-\eta)^2}\left(\frac{1}{\xi-\zeta}+\frac{1}{\zeta-\eta}\right)+\frac{1}{\xi-\eta}\frac{1}{(\xi-\zeta)^2},\\
&\frac{1}{(\xi-\zeta)^2}\frac{1}{(\zeta-\eta)^2}=\frac{2}{(\xi-\eta)^3}\left(\frac{1}{\xi-\zeta}+\frac{1}{\zeta-\eta}\right)+\frac{1}{(\xi-\eta)^2}\left[\frac{1}{(\xi-\zeta)^2}+\frac{1}{(\zeta-\eta)^2}\right],\\
      &\frac{1}{\xi-\zeta}\frac{1}{(\zeta-\eta)^3}=\frac{1}{(\xi-\eta)^3}\left(\frac{1}{\xi-\zeta}+\frac{1}{\zeta-\eta}\right)+\frac{1}{(\xi-\eta)^2}\frac{1}{(\zeta-\eta)^2}+\frac{1}{\xi-\eta}\frac{1}{(\zeta-\eta)^3},
    \end{aligned}
  \end{equation}
into  \eqref{eq:KsR12} and collecting  the  coefficients of $(\xi-\eta)^{-3}$, $(\xi-\eta)^{-2}$ and $(\xi-\eta)^{-1}$. The coefficient of  $(\xi-\eta)^{-3}$  vanishes identically. The $(\xi-\eta)^{-2}$ contribution is
 \begin{equation}
   \begin{aligned}
    & \sum_{\zeta\in\Omega}\left\{\left(\frac{1}{\xi-\zeta}+\frac{1}{\zeta-\eta}\right)\mathbf{f}_0^\T(\xi)\mathbf{g}_0(\zeta)\left[\mathbf{F}_1^\T(\zeta)\mathbf{G}_0(\eta)-\mathbf{F}_0^\T(\zeta)\mathbf{G}_1(\eta)\right]\right.\\
    & +\frac{1}{(\zeta-\eta)^2}\left[-2\mathbf{f}_0^\T(\xi)\mathbf{g}_0(\zeta)\mathbf{F}_0^\T(\zeta)\mathbf{G}_0(\eta)\right]\\
    &+\left(\frac{1}{\xi-\zeta}+\frac{1}{\zeta-\eta}\right)\left[\mathbf{f}_0^\T(\xi)\mathbf{g}_0(\zeta)\mathbf{F}_0^\T(\zeta)\mathbf{G}_1(\eta)+\mathbf{f}_0^\T(\xi)\mathbf{g}_1(\zeta)\mathbf{F}_0^\T(\zeta)\mathbf{G}_0(\eta)\right]\\
    &+\left.\left[\frac{1}{(\xi-\zeta)^2}+\frac{1}{(\zeta-\eta)^2}\right]\mathbf{f}_0^\T(\xi)\mathbf{g}_0(\zeta)\mathbf{F}_0^\T(\zeta)\mathbf{G}_0(\eta)\right\}\\
    =&\sum_{\zeta\in\Omega}\left\{\left[\frac{1}{(\xi-\zeta)^2}-\frac{1}{(\zeta-\eta)^2}\right]\mathbf{f}_0^\T(\xi)\mathbf{g}_0(\zeta)\mathbf{F}_0^\T(\zeta)\mathbf{G}_0(\eta)\right.\\
    &+\left.\left(\frac{1}{\xi-\zeta}+\frac{1}{\zeta-\eta}\right)\left[\mathbf{f}_0^\T(\xi)\mathbf{g}_0(\zeta)\mathbf{F}_1^\T(\zeta)\mathbf{G}_0(\eta)+\mathbf{f}_0^\T(\xi)\mathbf{g}_1(\zeta)\mathbf{F}_0^\T(\zeta)\mathbf{G}_0(\eta)\right]\right\}\\
    =&\left[\mathcal{K}\begin{pmatrix}
      \mathbf{F}_0^\T \\
      \mathbf{F}_1^\T
    \end{pmatrix}(\xi)\right]_{(1)}\mathbf{G}_0(\eta)+\mathbf{f}_0^\T(\xi)\left[\mathcal{R}^\T(
      \mathbf{g}_0,\mathbf{g}_1
    )(\eta)\right]_{[1]},
   \end{aligned}
 \end{equation}
  where the subscript $(j)$ denotes the $j$-th row of the matrix, while the subscript $[j]$ denotes the $j$-th column.
  Similarly, the $(\xi-\eta)^{-1}$ contribution is
  \begin{equation}
   \begin{aligned}
    & \sum_{\zeta\in\Omega}\left\{\left(\frac{1}{\xi-\zeta}+\frac{1}{\zeta-\eta}\right)\mathbf{f}_0^\T(\xi)\mathbf{g}_0(\zeta)\mathbf{F}_1^\T(\zeta)\mathbf{G}_1(\eta)\right.\\
    & +\frac{\mathbf{f}_0^\T(\xi)\mathbf{g}_0(\zeta)}{(\zeta-\eta)^2}\left[\mathbf{F}_1^\T(\zeta)\mathbf{G}_0(\eta)-\mathbf{F}_0^\T(\zeta)\mathbf{G}_1(\eta)\right]-\frac{2\mathbf{f}_0^\T(\xi)\mathbf{g}_0(\zeta)\mathbf{F}_0^\T(\zeta)\mathbf{G}_0(\eta)}{(\zeta-\eta)^3}\\
    &+\left(\frac{1}{\xi-\zeta}+\frac{1}{\zeta-\eta}\right)\left[\mathbf{f}_0^\T(\xi)\mathbf{g}_1(\zeta)\mathbf{F}_0^\T(\zeta)\mathbf{G}_1(\eta)\right]+\frac{\mathbf{f}_0^\T(\xi)\mathbf{g}_1(\zeta)\mathbf{F}_0^\T(\zeta)\mathbf{G}_0(\eta)}{(\zeta-\eta)^2}\\
    &+\left.\frac{\mathbf{f}_0^\T(\xi)\mathbf{g}_0(\zeta)\mathbf{F}_0^\T(\zeta)\mathbf{G}_1(\eta)}{(\xi-\zeta)^2}\right\}\\
    =&\left[\mathcal{K}\begin{pmatrix}
      \mathbf{F}_0^\T \\
      \mathbf{F}_1^\T
    \end{pmatrix}(\xi)\right]_{(1)}\mathbf{G}_1(\eta)+\mathbf{f}_0^\T(\xi)\left[\mathcal{R}^\T(
      \mathbf{g}_0,\mathbf{g}_1
    )(\eta)\right]_{[2]}.
   \end{aligned}
 \end{equation}
Combining these results with the identity \begin{equation}
  \mathbf{R}_{12}(\xi,\eta)-\mathbf{K}_{12}(\xi,\eta)=\frac{\mathbf{F}_0^\T(\xi)\mathbf{G}_1(\eta)-\mathbf{f}_0^\T(\xi)\mathbf{g}_1(\eta)}{\xi-\eta}+\frac{\mathbf{F}_0^\T(\xi)\mathbf{G}_0(\eta)-\mathbf{f}_0^\T(\xi)\mathbf{g}_0(\eta)}{(\xi-\eta)^2},
\end{equation} we obtain
 \begin{equation}\label{eq:KRKR}
 \begin{aligned}
      &(\mathbf{K}\circ\mathbf{R}+\mathbf{K}-\mathbf{R})_{12}(\xi,\eta)\\
     = &\frac{\mathbf{f}_0^\T(\xi)\left[(\mathbbm{1}+\mathcal{R}^\T)(
      \mathbf{g}_0,\mathbf{g}_1
    )(\eta)\right]_{[2]}-\left[(\mathbbm{1}-\mathcal{K})\begin{pmatrix}
      \mathbf{F}_0^\T \\
      \mathbf{F}_1^\T
    \end{pmatrix}(\xi)\right]_{(1)}\mathbf{G}_1(\eta)}{\xi-\eta}\\
      &+\frac{\mathbf{f}_0^\T(\xi)\left[(\mathbbm{1}+\mathcal{R}^\T)(
      \mathbf{g}_0,\mathbf{g}_1
    )(\eta)\right]_{[1]}-\left[(\mathbbm{1}-\mathcal{K})\begin{pmatrix}
      \mathbf{F}_0^\T \\
      \mathbf{F}_1^\T
    \end{pmatrix}(\xi)\right]_{(1)}\mathbf{G}_0(\eta)}{(\xi-\eta)^2}.
 \end{aligned}
 \end{equation}
With the definition   \eqref{FfGg2}, \begin{equation}
    \begin{pmatrix}
      \mathbf{f}_0^\T \\
      \mathbf{f}_1^\T
    \end{pmatrix}=(\mathbbm{1}-\mathcal{K})\begin{pmatrix}
      \mathbf{F}_0^\T \\
      \mathbf{F}_1^\T
    \end{pmatrix},\quad \begin{pmatrix}
      \mathbf{G}_0,\mathbf{G}_1
    \end{pmatrix}=(\mathbbm{1}+\mathcal{R}^\T)\begin{pmatrix}
      \mathbf{g}_0,\mathbf{g}_1
    \end{pmatrix}
\end{equation} every term on the right–hand side of \eqref{eq:KRKR} cancels, and the $(1,2)$-entry of \eqref{eq:KsR} is verified.  The remaining entries follow in the same manner.
  \end{proof}
\begin{problem}\label{RH3}
Find a matrix-valued function $\mathcal{M}(\xi)\in \mathbb{C}^{(n+1)\times(n+1)}$ with the following properties:
  \begin{enumerate}[(i)]
    \item   $\mathcal{M}(\xi)$ is analytic on $\mathbb{C}\backslash\Omega$.
    \item As $\xi\rightarrow\infty$, $\mathcal{M}(\xi)=\mathbf{I}+O(\xi^{-1})$.
    \item At every point $\kappa$ in $\Omega$, $\xi=\kappa$  is a second-order pole of $\mathcal{M}(\xi)$, and  the residues are generalized, satisfying
    \begin{equation}\label{res:gr1}
    \begin{aligned}
      &\underset{\xi=\kappa}{\res}(\xi-\kappa)\mathcal{M}(\xi)=\lim_{\xi\rightarrow \kappa}\mathcal{M}(\xi)\mathbf{w}_0(\kappa),\\
      &\underset{\xi=\kappa}{\res}\mathcal{M}(\xi)=\lim_{\xi\rightarrow \kappa}\left[\mathcal{M}(\xi)\mathbf{w}_1(\kappa)+\partial_\xi[\mathcal{M}(\xi)]\mathbf{w}_0(\kappa)\right],
    \end{aligned}
            \end{equation}
            where $\mathbf{w}_0(\kappa)=\mathbf{f}_0(\kappa)\mathbf{g}_0^\T(\kappa)$ and $\mathbf{w}_1(\kappa)=\mathbf{f}_1(\kappa)\mathbf{g}_0^\T(\kappa)+\mathbf{f}_0(\kappa)\mathbf{g}_1^\T(\kappa)$.
  \end{enumerate}
\end{problem}
\begin{lemma}\label{lem:Ma}
Let $\mathcal{M}(\xi)$ be a solution of RH problem \ref{RH3}. Then,  for every $\kappa\in\Omega$, the matrix-valued function
 \begin{equation}\label{eq:Mp}
   \mathcal{M}(\xi)\left[\mathbf{I}-\frac{\mathbf{w}_0(\kappa)}{(\xi-\kappa)^2}-\frac{\mathbf{w}_1(\kappa)}{\xi-\kappa}\right]
 \end{equation} is analytic in a neighborhood of $\kappa$.
\end{lemma}
\begin{proof}
 In a neighborhood around $\kappa$, suppose that the solution $\mathcal{M}(\xi)$ and  its derivative  $\partial_\xi[\mathcal{M}(\xi)]$ possesses   Laurent expansions
\begin{equation}
\begin{aligned}
 & \mathcal{M}(\xi)=\mathcal{M}_{-2}(\xi-\kappa)^{-2}+\mathcal{M}_{-1}(\xi-\kappa)^{-1}+\mathcal{M}_{0}+\mathcal{M}_{1}(\xi-\kappa)+O((\xi-\kappa)^2),\\
 & \partial_\xi[\mathcal{M}(\xi)]=-2\mathcal{M}_{-2}(\xi-\kappa)^{-3}-\mathcal{M}_{-1}(\xi-\kappa)^{-2}+\mathcal{M}_{1}+O(\xi-\kappa),
\end{aligned}
\end{equation}
with constant matrices $\mathcal{M}_{-2}$, $\mathcal{M}_{-1}$, $\mathcal{M}_{0}$, $\mathcal{M}_{1}$. Inserting these expansions into the product \eqref{eq:Mp},
collecting coefficients of $(\xi-\kappa)^j(-4\leqslant j\leqslant-1)$ yields
\begin{equation}\label{eq:cmp}
\begin{aligned}
  &  (\xi-\kappa)^{-4}:-\mathcal{M}_{-2}\mathbf{w}_0(\kappa),\\
&(\xi-\kappa)^{-3}:-\mathcal{M}_{-1}\mathbf{w}_0(\kappa)-\mathcal{M}_{-2}\mathbf{w}_1(\kappa),\\
&(\xi-\kappa)^{-2}:\mathcal{M}_{-2}-\mathcal{M}_{0}\mathbf{w}_0(\kappa)-\mathcal{M}_{-1}\mathbf{w}_1(\kappa),\\
&(\xi-\kappa)^{-1}:\mathcal{M}_{-1}-\mathcal{M}_{1}\mathbf{w}_0(\kappa)-\mathcal{M}_{0}\mathbf{w}_1(\kappa).
\end{aligned}
\end{equation}
From the orthogonality \eqref{eq:noncon} we have
\begin{equation}
  \mathbf{w}_0^2(\kappa)=\mathbf{0},\quad \mathbf{w}_1^2(\kappa)=\mathbf{0},\quad \mathbf{w}_0(\kappa)\mathbf{w}_1(\kappa)=\mathbf{0}.
\end{equation}
Consequently, the coefficient matrices for these powers in   \eqref{eq:cmp} vanish precisely when
\begin{equation}
\mathcal{M}_{-1}=\mathcal{M}_{1}\mathbf{w}_0(\kappa)+\mathcal{M}_{0}\mathbf{w}_1(\kappa),\quad  \mathcal{M}_{-2}=\mathcal{M}_{0}\mathbf{w}_0(\kappa),
\end{equation}
and these relations are exactly the residue conditions \eqref{res:gr1}. Thus the product is free of singularities at $\kappa$, hence analytic in a neighborhood of $\kappa$.
\end{proof}
\begin{lemma}\label{lem:un}
 If   RH problem \ref{RH3} admits a solution, then it is unique and satisfies
  \begin{equation}
    \det\left[\mathcal{M}(\xi)\right]\equiv 1.
  \end{equation}
\end{lemma}
\begin{proof}
Because $\left[\mathbf{w}_0(\kappa)+\mathbf{w}_1(\kappa)(\xi-\kappa)\right]^2=\mathbf{0}$, we have
\begin{equation}
  \det\left[\mathbf{I}+\frac{\mathbf{w}_0(\kappa)}{(\xi-\kappa)^2}+\frac{\mathbf{w}_1(\kappa)}{\xi-\kappa}\right]=\det\left[\mathbf{I}+\frac{\mathbf{w}_0(\kappa)+\mathbf{w}_1(\kappa)(\xi-\kappa)}{(\xi-\kappa)^2}\right]=1.
\end{equation} Consequently,
\begin{equation}
  \det\left[\mathcal{M}(\xi)\right]=\det\left[ \mathcal{M}(\xi)\left[\mathbf{I}-\frac{\mathbf{w}_0(\kappa)}{(\xi-\kappa)^2}-\frac{\mathbf{w}_1(\kappa)}{\xi-\kappa}\right]\right]
\end{equation}
is analytic near every $\kappa\in\Omega$. Since $\kappa\in\Omega$ is arbitrary, $\det\left[\mathcal{M}(\xi)\right]$ is entire. The asymptotic condition  $\mathcal{M}(\xi)\rightarrow\mathbf{I}$ forces $\det\left[\mathcal{M}(\xi)\right]\equiv 1$.

Now suppose $\mathcal{N}(\xi)$ is another solution of RH problem \ref{RH3}. Then
\begin{equation}
  \mathcal{N}(\xi)\mathcal{M}^{-1}(\xi)=\mathcal{N}(\xi)\left[\mathbf{I}-\frac{\mathbf{w}_0(\kappa)}{(\xi-\kappa)^2}-\frac{\mathbf{w}_1(\kappa)}{\xi-\kappa}\right]\left[\mathbf{I}-\frac{\mathbf{w}_0(\kappa)}{(\xi-\kappa)^2}-\frac{\mathbf{w}_1(\kappa)}{\xi-\kappa}\right]^{-1}\mathcal{M}^{-1}(\xi)
\end{equation}
has no singularities and is asymptotically equal to the identity matrix at infinity. By Liouville's theorem,   $\mathcal{M}(\xi)=\mathcal{N}(\xi)$.
\end{proof}
\begin{lemma}
  Let $\mathcal{M}(\xi)$ be the unique solution of RH Problem \ref{RH3}.
Then its inverse transpose $\mathcal{M}^{-\T}(\xi)$ solves the following RH problem:
  \begin{problem}\label{RH4}
Find a matrix-valued function $\mathcal{M}^{-\T}(\xi)\in\mathbb{C}^{(n+1)\times(n+1)}$ such that
  \begin{enumerate}[(i)]
    \item $\mathcal{M}^{-\T}(\xi)$ is analytic on $\mathbb{C}\backslash\Omega$.
    \item  $\mathcal{M}^{-\T}(\xi)=\mathbf{I}+O(\xi^{-1})$ as $\xi\rightarrow\infty$.
    \item Every $\kappa\in\Omega$  is a second-order pole of $\mathcal{M}^{-\T}(\xi)$  and the   generalized residues satisfy
    \begin{equation}\label{res:gr2}
    \begin{aligned}
      &\underset{\xi=\kappa}{\res}(\xi-\kappa)\mathcal{M}^{-\T}(\xi)=-\lim_{\xi\rightarrow \kappa}\mathcal{M}^{-\T}(\xi)\mathbf{w}_0^{\T}(\kappa),\\
      &\underset{\xi=\kappa}{\res}\mathcal{M}^{-\T}(\xi)=-\lim_{\xi\rightarrow \kappa}\left[\mathcal{M}^{-\T}(\xi)\mathbf{w}_1^{\T}(\kappa)+\partial_\xi\left[\mathcal{M}^{-\T}(\xi)\right]\mathbf{w}_0^{\T}(\kappa)\right].
    \end{aligned}
            \end{equation}
  \end{enumerate}
\end{problem}
\end{lemma}
\begin{proof}
 Following the proof of Lemma \ref{lem:Ma}, the product
 \begin{equation}
    \mathcal{M}^{-\T}(\xi)\left[\mathbf{I}+\frac{\mathbf{w}^\T_0(\kappa)}{(\xi-\kappa)^2}+\frac{\mathbf{w}_1^\T(\kappa)}{\xi-\kappa}\right]
 \end{equation}
  is analytic in a neighborhood of every $\kappa\in\Omega$. Consequently,\begin{equation}
   \mathcal{M}^{-\T}(\xi)\mathcal{M}^\T(\xi)=\mathcal{M}^{-\T}(\xi)\left[\mathbf{I}+\frac{\mathbf{w}^\T_0(\kappa)}{(\xi-\kappa)^2}+\frac{\mathbf{w}_1^\T(\kappa)}{\xi-\kappa}\right] \left[\mathbf{I}-\frac{\mathbf{w}_0(\kappa)}{(\xi-\kappa)^2}-\frac{\mathbf{w}_1(\kappa)}{\xi-\kappa}\right]^\T\mathcal{M}^\T(\xi)
 \end{equation}
 is entire.  Because  both $\mathcal{M}(\xi)$ and $\mathcal{M}^{-\T}(\xi)$ tend to $\mathbf{I}$ as $\xi\rightarrow\infty$, Liouville’s theorem gives $\mathcal{M}^{-\T}(\xi)\mathcal{M}^\T(\xi)\equiv \mathbf{I}$. Uniqueness of the solution to RH Problem \ref{RH3} (Lemma \ref{lem:un}) then guarantees that $\mathcal{M}^{-\T}(\xi)$ is indeed the inverse transpose of $\mathcal{M}(\xi)$.
\end{proof}
\begin{theorem}\label{TH:RHiff}
  The operator  $\mathbbm{1}-\mathcal{K}$ is invertible if and only if  RH problem \ref{RH3} admits a unique solution.
\end{theorem}
\begin{proof}
    Assume    $\mathbbm{1}-\mathcal{K}$ is invertible. Define
\begin{equation}\label{def:M5}
  \mathcal{M}(\xi)=\mathbf{I}+\sum_{\eta\in\Omega}\left[\frac{\mathbf{F}_0(\eta)\mathbf{g}_0^\T(\eta)}{(\xi-\eta)^2}+\frac{\mathbf{F}_1(\eta)\mathbf{g}_0^\T(\eta)+\mathbf{F}_0(\eta)\mathbf{g}_1^\T(\eta)}{\xi-\eta}\right],
\end{equation}
where  $\mathbf{F}_0(\eta)$ and $\mathbf{F}_1(\eta)$ are given in  \eqref{FfGg2}. Clearly $\mathcal{M}(\xi)$  is analytic off $\Omega$ and tends to $\mathbf{I}$ as $\xi\rightarrow\infty$. It remains to verify the residue conditions \eqref{res:gr1}.

 Right-multiplying \eqref{def:M5} by $\mathbf{f}_0(\kappa)$ and using the orthogonality \eqref{eq:noncon} yields
  \begin{equation}
     \mathcal{M}(\xi)\mathbf{f}_0(\kappa)=\mathbf{f}_0(\kappa)+\sum_{\substack{\eta\in\Omega\\\eta\neq \kappa}}\left[\frac{\mathbf{F}_0(\eta)\mathbf{g}_0^\T(\eta)\mathbf{f}_0(\kappa)}{(\xi-\eta)^2}+\frac{\mathbf{F}_1(\eta)\mathbf{g}_0^\T(\eta)\mathbf{f}_0(\kappa)+\mathbf{F}_0(\eta)\mathbf{g}_1^\T(\eta)\mathbf{f}_0(\kappa)}{\xi-\eta}\right].
  \end{equation}
  Differentiating \eqref{def:M5} with respect to $\xi$ and right-multiplying by $\mathbf{f}_0(\kappa)$, then adding   $\mathcal{M}(\xi)\mathbf{f}_1(\kappa)$, we obtain
  \begin{equation}
  \begin{aligned}
  &\mathcal{M}(\xi)\mathbf{f}_1(\kappa)+\partial_\xi[\mathcal{M}(\xi)]\mathbf{f}_0(\kappa)\\
  =&\mathbf{f}_1(\kappa)+\sum_{\substack{\eta\in\Omega\\\eta\neq \kappa}}\left[\frac{\mathbf{F}_0(\eta)\mathbf{g}_0^\T(\eta)\mathbf{f}_1(\kappa)}{(\xi-\eta)^2}+\frac{\mathbf{F}_1(\eta)\mathbf{g}_0^\T(\eta)\mathbf{f}_1(\kappa)+\mathbf{F}_0(\eta)\mathbf{g}_1^\T(\eta)\mathbf{f}_1(\kappa)}{\xi-\eta}\right]\\
    &-\sum_{\substack{\eta\in\Omega\\\eta\neq \kappa}}\left[\frac{2\mathbf{F}_0(\eta)\mathbf{g}_0^\T(\eta)\mathbf{f}_0(\kappa)}{(\xi-\eta)^3}+\frac{\mathbf{F}_1(\eta)\mathbf{g}_0^\T(\eta)\mathbf{f}_0(\kappa)+\mathbf{F}_0(\eta)\mathbf{g}_1^\T(\eta)\mathbf{f}_0(\kappa)}{(\xi-\eta)^2}\right].
  \end{aligned}
  \end{equation}
  Letting $\xi\rightarrow\kappa$  and involving the relationship \eqref{FfGg2} yields
 \begin{equation}\label{eq:FMf}
 \begin{aligned}
    & \lim_{\xi\rightarrow \kappa}\left[\mathcal{M}(\xi)\mathbf{f}_0(\kappa)\right]^\T=\mathbf{f}_0^\T(\kappa)+\left[\mathcal{K}\begin{pmatrix}
                                                                                                                 \mathbf{F}_0^\T \\
                                                                                                                 \mathbf{F}_0^\T
                                                                                                               \end{pmatrix}(\kappa)\right]_{(1)}=\mathbf{F}_0^\T(\kappa),\\
   &  \lim_{\xi\rightarrow \kappa}\left[\mathcal{M}(\xi)\mathbf{f}_1(\kappa)+\partial_\xi[\mathcal{M}(\xi)]\mathbf{f}_0(\kappa)\right]^\T=\mathbf{f}_1^\T(\kappa)+\left[\mathcal{K}\begin{pmatrix}
                                                                                                                 \mathbf{F}_0^\T \\
                                                                                                                 \mathbf{F}_0^\T
                                                                                                               \end{pmatrix}(\kappa)\right]_{(2)}=\mathbf{F}_1^\T(\kappa).
 \end{aligned}
  \end{equation}
 From \eqref{def:M5} we read off the residues
 \begin{equation}\label{eq:Mdefres}
    \begin{aligned}
      &\underset{\xi=\kappa}{\res}(\xi-\kappa)\mathcal{M}(\xi)=\mathbf{F}_0(\kappa)\mathbf{g}_0^\T(\kappa),\\
      &\underset{\xi=\kappa}{\res}\mathcal{M}(\xi)=\mathbf{F}_1(\kappa)\mathbf{g}_0^\T(\kappa)+\mathbf{F}_0(\kappa)\mathbf{g}_1^\T(\kappa),
    \end{aligned}
            \end{equation}
            which together with \eqref{eq:FMf} satisfy \eqref{res:gr1}.  Hence $\mathcal{M}(\xi)$ solves RH Problem \ref{RH3}.

 Viceversa, assume RH Problem \ref{RH3} has a unique solution  $ \mathcal{M}(\xi)$. Define
 \begin{alignat}{1}
   &\mathbf{F}_0(\kappa)=\lim_{\xi\rightarrow \kappa}  \mathcal{M}(\xi)\mathbf{f}_0(\kappa),\quad \mathbf{F}_1(\kappa)= \lim_{\xi\rightarrow \kappa}\left[\partial_\xi [\mathcal{M}(\xi)] \mathbf{f}_0(\kappa)+\mathcal{M}(\xi) \mathbf{f}_1(\kappa)\right],\label{eq:FGMr}\\
   &\mathbf{G}_0(\kappa)=\lim_{\xi\rightarrow \kappa}  \mathcal{M}^{-\T}(\xi)\mathbf{g}_0(\kappa),\nonumber\\
   &\mathbf{G}_1(\kappa)= \lim_{\xi\rightarrow \kappa}\left[\mathcal{M}^{-\T}(\xi) \mathbf{g}_1(\kappa)-\mathcal{M}^{-\T}(\xi)\partial_\xi [\mathcal{M}^\T(\xi)]\mathcal{M}^{-\T}(\xi) \mathbf{g}_0(\kappa)\right],\label{eq:FGMr1}
 \end{alignat}
 where $\mathcal{M}^{-\T}(\xi)$ is the inverse transpose of $\mathcal{M}(\xi)$. Conversely,
 \begin{equation}\label{eq:FGMr2}
   \mathbf{g}_0(\kappa)=\lim_{\xi\rightarrow \kappa}  \mathcal{M}^{\T}(\xi)\mathbf{G}_0(\kappa),\quad \mathbf{g}_1(\kappa)= \lim_{\xi\rightarrow \kappa}\left[\mathcal{M}^{\T}(\xi) \mathbf{G}_1(\kappa)+\partial_\xi [\mathcal{M}^\T(\xi)]\mathbf{G}_0(\kappa)\right].
 \end{equation} The solutions of RH problems \ref{RH3} and \ref{RH4} can be expressed by
 \begin{alignat}{1}
  &\mathcal{M}(\xi)=\mathbf{I}+\sum_{\eta\in\Omega}\left[\frac{\mathbf{F}_0(\eta)\mathbf{g}_0^\T(\eta)}{(\xi-\eta)^2}+\frac{\mathbf{F}_1(\eta)\mathbf{g}_0^\T(\eta)+\mathbf{F}_0(\eta)\mathbf{g}_1^\T(\eta)}{\xi-\eta}\right],\label{eq:Mitd1}\\
  & \mathcal{M}^{-\T}(\xi)=\mathbf{I}-\sum_{\eta\in\Omega}\left[\frac{\mathbf{G}_0(\eta)\mathbf{f}_0^\T(\eta)}{(\xi-\eta)^2}+\frac{\mathbf{G}_1(\eta)\mathbf{f}_0^\T(\eta)+\mathbf{G}_0(\eta)\mathbf{f}_1^\T(\eta)}{\xi-\eta}\right].\label{eq:Mitd2}
 \end{alignat}
Inserting \eqref{eq:Mitd1} into \eqref{eq:FGMr}–\eqref{eq:FGMr2} yields the coupled equations
\begin{alignat}{1}
&\begin{pmatrix}
      \mathbf{F}_0^\T(\xi) \\
      \mathbf{F}_1^\T(\xi)
    \end{pmatrix}=\begin{pmatrix}
      \mathbf{f}_0^\T(\xi) \\
      \mathbf{f}_1^\T(\xi)
    \end{pmatrix}+\mathcal{K}\begin{pmatrix}
      \mathbf{F}_0^\T \\
      \mathbf{F}_1^\T
    \end{pmatrix}(\xi),\\
    &(
      \mathbf{g}_0(\xi),\mathbf{g}_1(\xi)
 )= (
      \mathbf{G}_0(\xi),\mathbf{G}_1(\xi)
    )-\mathcal{R}^\T(
      \mathbf{g}_0,\mathbf{g}_1
 )(\xi).
\end{alignat}
A repeat of the argument in Theorem \ref{TH:th1} shows that $ \mathcal{K}\circ\mathcal{R}=\mathcal{R}-\mathcal{K}$, where the operator $
  \mathcal{R}$  is the resolvent operator with kernel \eqref{def:R1} and with $\mathbf{F}_0$, $\mathbf{F}_1$, $\mathbf{G}_0$, $\mathbf{G}_1$ defined by \eqref{eq:FGMr}.  Hence  $\mathbbm{1}-\mathcal{K}$ is invertible.
\end{proof}
  \end{theorem}
 \section{Connection to discrete orthogonal polynomials}
 Discrete orthogonal polynomials are a class of polynomial sequences \(\{P_n(x)\}_{n=0}^{\infty}\) that satisfy an orthogonality condition with respect to a  discrete weight function  over a discrete set of points.
\begin{definition}\label{def:dops}
  Let $\mathcal{X}$ be a discrete set  of real or complex numbers, and let $w(x) > 0$ be a weight function defined on $\mathcal{X}$. A sequence of monic polynomials $\{P_n(x)\}$  is said to be orthogonal if
\begin{equation}
  \sum_{x \in \mathcal{X}} P_m(x) P_n(x) w(x) = h_n \delta_{mn},
\end{equation}
where $\delta_{mn}$ is the Kronecker delta, and $h_n > 0$ are normalization constants.
\end{definition}

 All discrete orthogonal polynomials  satisfy a linear three-term recurrence relation. For monic discrete orthogonal polynomials $\{P_n(x)\}$, this recurrence takes the form
\begin{equation}
  P_{n+1}(x) = (x - \alpha_n) P_n(x) - \beta_n P_{n-1}(x), \quad \text{for } n \geqslant 0,
\end{equation}
with initial conditions:
\begin{equation}
  P_{-1}(x) = 0, \quad P_0(x) = 1,
\end{equation}
where  the recurrence coefficients $\alpha_n$ and $\beta_n$ are  given by 
  \begin{equation}
      \alpha_n = \frac{1}{h_n} \sum_{x \in \mathcal{X}} x P_n^2(x) w(x),\quad \beta_n = \frac{h_n}{h_{n-1}} \quad (\beta_0 = 0),
  \end{equation}
 and $h_n = \sum_{x \in \mathcal{X}} P_n^2(x) w(x)$ is the squared norm of $P_n(x)$.
 \begin{definition}
  Given a sequence of discrete orthogonal polynomials $\{P_n(x)\}$ with squared norms $\{h_n\}$, the Darboux--Christoffel kernel of order $N$ is defined as:
\begin{equation}
  K_N(x, y) = \sqrt{w(x)w(y)}\sum_{n=0}^N \frac{P_n(x) P_n(y)}{h_n}.
\end{equation}
\end{definition}

This kernel satisfies the reproducing property
\begin{equation}
  \sum_{y\in \mathcal{X}} K_N(x, y)K_N(y, z) =K_N(x, z).
\end{equation}
A key identity for the Darboux--Christoffel kernel is the Christoffel--Darboux formula
\begin{equation}\label{eq:DCid}
   K_N(x, y) = \sqrt{w(x)w(y)}\frac{P_{N+1}(x) P_N(y) - P_N(x) P_{N+1}(y)}{h_N(x - y)}, \quad x\neq y, 
\end{equation}
and for $x=y$,
\begin{equation}
  K_N(x,x)=\frac{w(x)}{h_N}\left[P_{N+1}'(x)P_N(x)-P_N'(x)P_{N+1}(x)\right].
\end{equation}

\begin{definition}
  We define the  $2\times 2$ matrix kernel
\begin{equation}
  \mathbf{K}_N(\xi,\eta)=\sum_{n=0}^N \frac{\sqrt{w(\xi)w(\eta)}}{h_n}\begin{pmatrix}
                                                     P_n(\xi) \\
                                                     P_n'(\xi) 
                                                   \end{pmatrix}\left(P_n(\eta),P_n'(\eta)\right),
\end{equation}
where $\{P_n(\xi)\}$ are the discrete orthogonal polynomials as defined in Definition  \ref{def:dops}.
\end{definition}
We immediately observe the following properties:
\begin{enumerate}[(i)]
  \item $\mathbf{K}_N(\xi,\eta)=\mathbf{K}_N^{\T}(\eta,\xi)$;
  \item $\displaystyle\sum_{\eta\in \mathcal{X}} \mathbf{K}_N(\xi, \eta)\begin{pmatrix}
                                                     0&1 \\
                                                     1&0
                                                   \end{pmatrix}\mathbf{K}_N(\eta, \zeta) =\sum_{n=1}^N \frac{n\sqrt{w(\xi)w(\zeta)}}{h_n}\begin{pmatrix}
                                                     P_n(\xi) &P_{n-1}(\xi)\\
                                                     P_n'(\xi) &P_{n-1}'(\xi)
                                                   \end{pmatrix}\begin{pmatrix}
                                                     P_{n-1}(\zeta) &P_{n-1}'(\zeta)\\
                                                     P_n(\zeta) &P_n'(\zeta)
                                                   \end{pmatrix}$.
\end{enumerate}

Let \begin{equation}
\mathbf{f}_0(\xi)=\frac{\sqrt{w(\xi)}}{h_N}\begin{pmatrix}
  P_{N+1}(\xi)\\P_N(\xi)
\end{pmatrix}, \quad \mathbf{g}_0(\eta)=\sqrt{w(\eta)}\begin{pmatrix}
                                          P_N(\eta) \\
                                          -P_{N+1}(\eta) 
                                        \end{pmatrix},
\end{equation}
and 
\begin{equation}
\mathbf{f}_1(\xi)=\frac{\sqrt{w(\xi)}}{h_N}\begin{pmatrix}
  P_{N+1}'(\xi)\\P_N'(\xi)
\end{pmatrix}, \quad \mathbf{g}_1(\eta)=\sqrt{w(\eta)}\begin{pmatrix}
                                          P_N'(\eta) \\
                                          -P_{N+1}'(\eta) 
                                        \end{pmatrix}.
\end{equation}
Using the Christoffel--Darboux identity \eqref{eq:DCid}, we obtain:
\begin{equation}
  \mathbf{K}_N(\xi,\eta)=
    \begin{pmatrix}
                            \frac{\mathbf{f}_0^\T(\xi)\mathbf{g}_0(\eta)}{\xi-\eta} & \frac{\mathbf{f}_0^\T(\xi)\mathbf{g}_1(\eta)}{\xi-\eta}+\frac{\mathbf{f}_0^\T(\xi)\mathbf{g}_0(\eta)}{(\xi-\eta)^2} \\
                            \frac{\mathbf{f}_1^\T(\xi)\mathbf{g}_0(\eta)}{\xi-\eta}-\frac{\mathbf{f}_0^\T(\xi)\mathbf{g}_0(\eta)}{(\xi-\eta)^2} &  \frac{\mathbf{f}_1^\T(\xi)\mathbf{g}_1(\eta)}{\xi-\eta}+ \frac{\mathbf{f}_1^\T(\xi)\mathbf{g}_0(\eta)-\mathbf{f}_0^\T(\xi)\mathbf{g}_1(\eta)}{(\xi-\eta)^2}-\frac{2\mathbf{f}_0^\T(\xi)\mathbf{g}_0(\eta)}{(\xi-\eta)^3}
                          \end{pmatrix}.
\end{equation}
which matches the structure given in \eqref{eq:mkk}.
  \section{Generalized discrete integrable operator}
The original discrete integrable operator is inherently connected to the RH problem for pure soliton solutions, with its transmission coefficients showing only simple poles on the spectral plane. Does a specific discrete integrable operator exist whose construction can precisely match the properties of higher-order pole solutions? The transmission coefficients corresponding to these solutions exhibit higher-order singularities.  Inspired by this, we successfully built a discrete integrable operator that perfectly aligns with the analytic and algebraic nature of higher-order pole data by studying their core analytic and algebraic properties.

    Let $\Omega$ be a discrete set and let $\mathfrak{m}_\xi\in\mathbb{N}$  be a multiplicity attached to every $\xi\in\Omega$.
 Given differentiable column vector functions
$\mathbf{f}, \mathbf{g}:\mathbb{C}\rightarrow\mathbb{C}^{n+1} $
  satisfying the orthogonality relations
  \begin{equation}\label{re:uv}
  \partial_\xi^{j}\left[\mathbf{f}^\T(\xi)\right]\partial_\xi^{k}\left[\mathbf{g}(\xi)\right] =0,\quad 0\leqslant j,k\leqslant\mathfrak{m}_\xi+1,\quad \xi\in\Omega.
\end{equation}
We use $\mathbb{C}^{n+1}$ rather than $\mathbb{C}^n$ so that the $n$-component integrable hierarchy can be handled more conveniently in Section \ref{sec:ihho}.
\begin{definition}
  We introduce discrete integral type operators $\mathscr{K}$ and $\mathscr{K}^\T$  acting on differentiable row  and column vector functions, respectively, as follows. For every differentiable column vector function
$\mathbf{h}:\mathbb{C}\rightarrow\mathbb{C}^{m} $,
    \begin{subequations}\label{def:gdio}
\begin{equation}\label{def:Krow}
(\mathscr{K}\mathbf{h}^\T)(\xi)=\sum_{\eta\in\Omega}\frac{1}{\mathfrak{m}_\eta!}\partial_\eta^{\mathfrak{m}_\eta}\left[\frac{\mathbf{f}^\T(\xi)\mathbf{g}(\eta)\mathbf{h}^\T(\eta)}{\xi-\eta}\right],
\end{equation}
\begin{equation}\label{def:Kcol}
(\mathscr{K}^\T\mathbf{h})(\eta)=\sum_{\xi\in\Omega}\frac{1}{\mathfrak{m}_\xi!}\partial_\xi^{\mathfrak{m}_\xi}\left[\frac{\mathbf{h}(\xi)\mathbf{f}^\T(\xi)\mathbf{g}(\eta)}{\xi-\eta}\right].
\end{equation}
\end{subequations}
\end{definition}

 Thanks to the orthogonality condition \eqref{re:uv}, the singular terms $\eta=\xi$ in \eqref{def:Krow} and $\xi=\eta$ in \eqref{def:Kcol} vanish identically.  Consequently, the sums may be restricted to distinct indices:
\begin{subequations}
\begin{align}
  &(\mathscr{K}\mathbf{h}^\T)(\xi)=\sum_{\substack{\eta\in\Omega\\\eta\neq\xi}}\frac{1}{\mathfrak{m}_\eta!}\partial_\eta^{\mathfrak{m}_\eta}\left[\frac{\mathbf{f}^\T(\xi)\mathbf{g}(\eta)\mathbf{h}^\T(\eta)}{\xi-\eta}\right],\\
  &(\mathscr{K}^\T\mathbf{h})(\eta)=\sum_{\substack{\xi\in\Omega\\\xi\neq\eta}}\frac{1}{\mathfrak{m}_\xi!}\partial_\xi^{\mathfrak{m}_\xi}\left[\frac{\mathbf{h}(\xi)\mathbf{f}^\T(\xi)\mathbf{g}(\eta)}{\xi-\eta}\right].
\end{align}
\end{subequations}
\begin{theorem}\label{TH:GRS}
 Assume the operator  $\mathbbm{1}-\mathscr{K}$ is invertible. Then its resolvent  $
  \mathscr{R}=(\mathbbm{1}-\mathscr{K})^{-1}-\mathbbm{1}$ can be represented as
\begin{subequations}\label{def:scrR}
\begin{align}
  &(\mathscr{R}\mathbf{h}^\T)(\xi)=\sum_{\eta\in\Omega}\frac{1}{\mathfrak{m}_\eta!}\partial_\eta^{\mathfrak{m}_\eta}\left[\frac{\mathbf{F}^\T(\xi)\mathbf{G}(\eta)\mathbf{h}^\T(\eta)}{\xi-\eta}\right],\\
  &(\mathscr{R}^\T\mathbf{h})(\eta)=\sum_{\xi\in\Omega}\frac{1}{\mathfrak{m}_\xi!}\partial_\xi^{\mathfrak{m}_\xi}\left[\frac{\mathbf{h}(\xi)\mathbf{F}^\T(\xi)\mathbf{G}(\eta)}{\xi-\eta}\right],
\end{align}
\end{subequations}
where the column-vector functions $\mathbf{F}$ and $\mathbf{G}$ are defined by
\begin{subequations}
\begin{align}
 & \mathbf{F}^\T(\xi)=[(\mathbbm{1}-\mathscr{K})^{-1}\mathbf{f}^\T](\xi),&& \partial_\xi^j[\mathbf{F}^\T(\xi)]=\partial_\xi^j[\mathbf{f}^\T(\xi)]+(\mathscr{K}_j\mathbf{F}^\T)(\xi),\label{reUu}\\
  &\mathbf{G}(\eta)=[(\mathbbm{1}-\mathscr{K}^\T)^{-1}\mathbf{g}](\eta),&& \partial_\eta^l[\mathbf{G}(\eta)]=\partial_\eta^l[\mathbf{g}(\eta)]+(\mathscr{K}^\T_l\mathbf{G})(\eta),\label{reVv}
\end{align}
\end{subequations}
and \begin{subequations}\label{eq:Kjdef}
\begin{align}
  &(\mathscr{K}_j\mathbf{h}^\T)(\xi)=\sum_{\substack{\eta\in\Omega\\\eta\neq\xi}}\frac{1}{\mathfrak{m}_\eta!}\partial_\xi^j\partial_\eta^{\mathfrak{m}_\eta}\left[\frac{\mathbf{f}^\T(\xi)\mathbf{g}(\eta)\mathbf{h}^\T(\eta)}{\xi-\eta}\right],\quad 1\leqslant j\leqslant\mathfrak{m}_\xi,\\
  &(\mathscr{K}_l^\T\mathbf{h})(\eta)=\sum_{\substack{\xi\in\Omega\\\xi\neq\eta}}\frac{1}{\mathfrak{m}_\xi!}\partial_\eta^l\partial_\xi^{\mathfrak{m}_\xi}\left[\frac{\mathbf{h}(\xi)\mathbf{f}^\T(\xi)\mathbf{g}(\eta)}{\xi-\eta}\right],\quad 1\leqslant l\leqslant\mathfrak{m}_\eta.
\end{align}
\end{subequations}
\end{theorem}
\begin{proof}
  The invertibility of $\mathbbm{1}-\mathscr{K}$ ensures the existence of a unique resolvent $\tilde{\mathscr{R}}$ satisfying
\begin{equation}
  (\mathbbm{1}-\mathscr{K})\circ (\mathbbm{1}+\tilde{\mathscr{R}})=(\mathbbm{1}+\tilde{\mathscr{R}})\circ (\mathbbm{1}-\mathscr{K})=\mathbbm{1}.
\end{equation}
By virtue of \eqref{reUu} and \eqref{reVv},
\begin{equation}\label{reUuVv1}
    \tilde{\mathscr{R}}\mathbf{f}^\T=\mathbf{F}^\T-\mathbf{f}^\T,\quad {\mathscr{K}}^\T\mathbf{G}=\mathbf{G}-\mathbf{g},\quad (\mathscr{K}^\T_l\mathbf{G})(\eta)= \partial_\eta^l[\mathbf{G}(\eta)]-\partial_\eta^l[\mathbf{g}(\eta)].
\end{equation}
Let $ \mathscr{R}$ be given by \eqref{def:scrR}. We must show $\mathscr{R}\circ\mathscr{K}=\mathscr{R}-\mathscr{K}$.
   Indeed,
  \begin{equation}
  \begin{aligned}
        &(\mathscr{R}\circ\mathscr{K})(\mathbf{h}^\T)(\xi)=\sum_{\zeta\in\Omega}\sum_{\eta\in\Omega}\frac{1}{\mathfrak{m}_\eta!\mathfrak{m}_\zeta!}\partial_\eta^{\mathfrak{m}_\eta}\partial_\zeta^{\mathfrak{m}_\zeta}\left[\frac{\mathbf{F}^\T(\xi)\mathbf{G}(\zeta)}{\xi-\zeta}\frac{\mathbf{f}^\T(\zeta)\mathbf{g}(\eta)\mathbf{h}^\T(\eta)}{\zeta-\eta}\right]\\
        =&\sum_{\eta\in\Omega}\frac{1}{\mathfrak{m}_\eta!}\partial_\eta^{\mathfrak{m}_\eta}\left[\frac{1}{\xi-\eta}\sum_{\zeta\in\Omega}\frac{1}{\mathfrak{m}_\zeta!}\partial_\zeta^{\mathfrak{m}_\zeta}\left[\mathbf{F}^\T(\xi)\mathbf{G}(\zeta)\mathbf{f}^\T(\zeta)\mathbf{g}(\eta)\mathbf{h}^\T(\eta)\left[\frac{1}{\xi-\zeta}+\frac{1}{\zeta-\eta}\right]\right]\right]\\
        =&\sum_{\eta\in\Omega}\frac{1}{\mathfrak{m}_\eta!}\partial_\eta^{\mathfrak{m}_\eta}\left[\frac{(\mathscr{R}\mathbf{f}^\T)(\xi)\mathbf{g}(\eta)+\mathbf{F}^\T(\xi)(\mathscr{K}^\T\mathbf{G})(\eta)}{\xi-\eta}\mathbf{h}^\T(\eta)\right]\\
        =&\sum_{\eta\in\Omega}\frac{1}{\mathfrak{m}_\eta!}\partial_\eta^{\mathfrak{m}_\eta}\left[\frac{[\mathbf{F}^\T(\xi)-\mathbf{f}^\T(\xi)]\mathbf{g}(\eta)+\mathbf{F}^\T(\xi)[\mathbf{G}(\eta)-\mathbf{g}(\eta)]}{\xi-\eta}\mathbf{h}^\T(\eta)\right]\\
        &+\sum_{\eta\in\Omega}\frac{1}{\mathfrak{m}_\eta!}\partial_\eta^{\mathfrak{m}_\eta}\left[\frac{\mathbf{F}^\T(\xi)(\mathscr{R}^\T\mathbf{G}-\tilde{\mathscr{R}}^\T\mathbf{G})(\eta)}{\xi-\eta}\mathbf{h}^\T(\eta)\right]\\
        =&\mathscr{R}(\mathbf{h}^\T)(\xi)-\mathscr{K}(\mathbf{h}^\T)(\xi)+\sum_{\eta\in\Omega}\frac{1}{\mathfrak{m}_\eta!}\partial_\eta^{\mathfrak{m}_\eta}\left[\frac{(\mathscr{R}\mathbf{f}^\T-\tilde{\mathscr{R}}\mathbf{f}^\T)(\xi)\mathbf{g}(\eta)\mathbf{h}^\T(\eta)}{\xi-\eta}\right].
  \end{aligned}
  \end{equation}
    The second-to-last equality in the above  follows from  \eqref{reUuVv1}. By uniqueness of the resolvent we conclude $\mathscr{R}\equiv\tilde{\mathscr{R}}$.
\end{proof}
\begin{problem}\label{RH2}
Find a matrix-valued function $\mathscr{M}(\xi)$ such that
  \begin{enumerate}[(i)]
    \item $\mathscr{M}(\xi)$ is analytic on $\mathbb{C}\backslash\Omega$.
    \item As $\xi\rightarrow\infty$, $\mathscr{M}(\xi)=\mathbf{I}+O(\xi^{-1})$.
    \item Every point $\tau$ in $\Omega$  is a $\mathfrak{m}_\tau+1$-order pole for $\mathscr{M}(\xi)$, and the generalized residue conditions
    \begin{equation}\label{res:gr}
      \underset{\xi=\tau}{\res}(\xi-\tau)^{\mathfrak{m}_\tau-\mathfrak{n}_\tau}\mathscr{M}(\xi)=\lim_{\xi\rightarrow\tau}\sum_{\substack{j,l\geqslant0\\j+l= \mathfrak{n}_\tau}}\frac{1}{j!l!}\mathscr{M}^{(j)}(\xi)\mathbf{w}^{(l)}(\tau),\quad
      0\leqslant \mathfrak{n}_\tau\leqslant\mathfrak{m}_{\tau},
            \end{equation}
            hold, where $\displaystyle\mathscr{M}^{(j)}(\xi)= \partial_\xi^{j}\left[\mathscr{M}(\xi)\right]$ and $ \mathbf{w}^{(l)}(\xi)= \partial_\xi^{l}\left[\mathbf{f}(\xi)\mathbf{g}^\T(\xi)\right]$.
  \end{enumerate}
\end{problem}
  \begin{remark}\label{remark:a}
   The orthogonality relation \eqref{re:uv} implies
\begin{equation}\label{re:ww}
  \mathbf{w}^{(j)}\mathbf{w}^{(l)}=\mathbf{0},\quad 0\leqslant j,l\leqslant \mathfrak{m}_\tau+1.
\end{equation}
 Assume the Laurent expansion of $\mathscr{M}(\xi)$ near $\tau$ is
\begin{equation}
\mathscr{M}(\xi)=\sum_{j=-\mathfrak{m}_\tau-1}^{\mathfrak{m}_\tau}\mathscr{M}_{j}(\xi-\tau)^{j}+O((\xi-\tau)^{\mathfrak{m}_\tau+1}),\quad \xi\rightarrow\tau.
\end{equation}
Then
\begin{equation}
\begin{aligned}
   & \mathscr{M}(\xi)\left[\mathbf{I}-\frac{\sum_{j=0}^{\mathfrak{m}_\tau}\frac{\mathbf{w}^{(j)}(\tau)}{j!}(\xi-\tau)^j}{(\xi-\tau)^{\mathfrak{m}_\tau+1}}\right]\\
    =&\sum_{j=-\mathfrak{m}_\tau-1}^{-1}\left[\mathscr{M}_j-\sum_{l=j+1}^{j+\mathfrak{m}_\tau+1}\frac{\mathscr{M}_l\mathbf{w}^{(j+\mathfrak{m}_\tau+1-l)}(\tau)}{(j+\mathfrak{m}_\tau+1-l)!}\right](\xi-\tau)^j\\
& -\sum_{s=-\mathfrak{m}_\tau-1}^{-1}\sum_{j=-\mathfrak{m}_\tau-1}^s\frac{\mathscr{M}_j\mathbf{w}^{(s-j)}(\tau)}{(s-j)!}(\xi-\tau)^{s-\mathfrak{m}_\tau-1}+O(1),\quad \xi\rightarrow\tau.
\end{aligned}
\end{equation}
Consequently, the product on the left is analytic  at $\xi=\tau$ if and only if
\begin{align}
    &\mathscr{M}_j=\sum_{l=j+1}^{j+\mathfrak{m}_\tau+1}\frac{\mathscr{M}_l\mathbf{w}^{(j+\mathfrak{m}_\tau+1-l)}(\tau)}{(j+\mathfrak{m}_\tau+1-l)!},\label{re:M1} \quad -\mathfrak{m}_{\tau}-1\leqslant j\leqslant-1,\\
  &\sum_{j=-\mathfrak{m}_\tau-1}^s\frac{\mathscr{M}_j\mathbf{w}^{(s-j)}(\tau)}{(s-j)!}=\mathbf{0},\quad -\mathfrak{m}_\tau-1\leqslant s\leqslant -1.\label{re:M2}
\end{align}
Indeed, \eqref{re:M1} is a direct consequence of the residue conditions \eqref{res:gr}, whereas \eqref{re:M2} follows from \eqref{re:ww} and \eqref{re:M1}.
  \end{remark}

  On the other hand, the identity
  \begin{equation}
    \left[\sum_{j=0}^{\mathfrak{m}_\tau}\frac{\mathbf{w}^{(j)}}{j!}(\xi-\tau)^j\right]^2=\mathbf{0}
  \end{equation} implies
\begin{equation}
  \det\left[\mathbf{I}-\frac{1}{(\xi-\tau)^{\mathfrak{m}_\tau+1}}\sum_{j=0}^{\mathfrak{m}_\tau}\frac{\mathbf{w}^{(j)}}{j!}(\xi-\tau)^j\right]\equiv1.
\end{equation}
Coupled with RH problem \ref{RH2} and Remark \ref{remark:a}, this shows that $\det [\mathscr{M}(\xi)]$ is  entire. Because $\det [\mathscr{M}(\xi)]\rightarrow 1$
as $\xi\rightarrow\infty$, Liouville's Theorem forces
\begin{equation}
  \det \mathscr{M}(\xi)\equiv 1,
\end{equation} which in particular guarantees the uniqueness of the solution (if it exists).  The   inverse transpose of $\mathscr{M}(\xi)$   satisfies the following RH problem:
 \begin{problem}\label{RH5}
Find a matrix-valued function $\mathscr{M}^{-\T}(\xi)$ such that
  \begin{enumerate}[(i)]
    \item $\mathscr{M}^{-\T}(\xi)$ is analytic on $\mathbb{C}\backslash\Omega$.
    \item As $\xi\rightarrow\infty$, $\mathscr{M}^{-\T}(\xi)=\mathbf{I}+O(\xi^{-1})$.
    \item Every $\tau$ in $\Omega$ is a $\mathfrak{m}_\tau+1$-order pole for $\mathscr{M}^{-\T}(\xi)$, and for $
     0\leqslant \mathfrak{n}_\tau\leqslant\mathfrak{m}_{\tau}$, the generalized residue conditions
    \begin{equation}
      \underset{\xi=\tau}{\res}(\xi-\tau)^{\mathfrak{m}_\tau-\mathfrak{n}_\tau}\mathscr{M}^{-\T}(\xi)=-\lim_{\xi\rightarrow\tau}\sum_{\substack{j,l\geqslant0\\j+l= \mathfrak{n}_\tau}}\frac{1}{j!l!}\left[\mathscr{M}^{-\T}\right]^{(j)}(\xi)\left[\mathbf{w}^{\T}\right]^{(l)}(\tau),
            \end{equation}hold,
            where $\displaystyle \left[\mathscr{M}^{-\T}\right]^{(j)}(\xi)= \partial_\xi^{j}\left[\mathscr{M}^{-\T}(\xi)\right], \left[\mathbf{w}^{\T}\right]^{(l)}(\xi)= \partial_\xi^{l}\left[\mathbf{g}(\xi)\mathbf{f}^\T(\xi)\right]$.
  \end{enumerate}
\end{problem}
\begin{theorem}\label{TH:GRHiif}
  The operator  $\mathbbm{1}-\mathscr{K}$ is invertible if and only if RH problem \ref{RH2} possesses  a unique solution.
\end{theorem}
\begin{proof}
    Assume  $\mathbbm{1}-\mathscr{K}$ is invertible. Define
\begin{equation}\label{def:M3}
  \mathscr{M}(\xi)=\mathbf{I}+\sum_{\eta\in\Omega}\frac{1}{\mathfrak{m}_\eta!}\partial_\eta^{\mathfrak{m}_\eta}\left[\frac{\mathbf{F}(\eta)\mathbf{g}^\T(\eta)}{\xi-\eta}\right],\quad\xi\in\Omega,
\end{equation}
where the column vector function  $\mathbf{F}(\eta)$ and its derivatives $\partial_\eta^j[\mathbf{F}(\eta)]$ are determined by  \eqref{reUu}.
As $\tau\in\Omega$, multiplying  \eqref{def:M3} on the right by $\mathbf{f}(\tau)$, and then taking the limit $\xi\rightarrow\tau$ yields
\begin{equation}
 \lim_{\xi\rightarrow\tau} [\mathscr{M}(\xi)\mathbf{f}(\tau)]^\T=\mathbf{f}^\T(\tau)+\sum_{\substack{\eta\in\Omega\\\eta\neq\tau}}\frac{1}{\mathfrak{m}_\eta!}\partial_\eta^{\mathfrak{m}_\eta}\left[\frac{\mathbf{f}^\T(\tau)\mathbf{g}(\eta)\mathbf{F}^\T(\eta)}{\tau-\eta}\right]=\mathbf{f}^\T(\tau)+(\mathscr{K}\mathbf{F}^\T)(\tau).
\end{equation}
For $1\leqslant m\leqslant\mathfrak{m}_\tau$, a similar computation gives
\begin{equation}
  \lim_{\xi\rightarrow\tau}\sum_{\substack{j,l\geqslant 0\\j+l=m}}\frac{m!}{j!l!}\mathscr{M}^{(j)}(\xi)\mathbf{f}^{(l)}(\tau)=\mathbf{f}^{(j)}(\tau)+\left[(\mathscr{K}_j\mathbf{F}^\T)(\tau)
  \right]^{\mathrm T},
\end{equation}
with the obvious notation
\begin{equation}
  \mathbf{f}^{(j)}(\xi)=\partial_\xi^j[\mathbf{f}(\xi)], \quad \mathbf{F}^{(j)}(\xi)=\partial_\xi^j[\mathbf{F}(\xi)],\quad\mathscr{M}^{(j)}(\xi)=\partial_\xi^j[\mathscr{M}(\xi)].
\end{equation}
Appealing to \eqref{reUu}  we obtain
\begin{equation}\label{reUM}
  \mathbf{F}^{(m)}(\tau)=\lim_{\xi\rightarrow\tau} \sum_{\substack{j,l\geqslant 0\\j+l=m}}\frac{m!}{j!l!}\mathscr{M}^{(j)}(\xi)\mathbf{f}^{(l)}(\tau),\quad 0\leqslant m\leqslant\mathfrak{m}_\tau.
\end{equation}
Inserting this identity into  \eqref{def:M3}  shows that $\mathscr{M}(\xi)$ satisfies the generalized residue conditions \eqref{res:gr}  as well as the remaining requirements  in RH problem \ref{RH2}.

Viceversa,  let $ \mathscr{M}(\xi)$  be the unique solution of  RH problem \ref{RH2}. Define the vectors
\begin{alignat}{1}
 & \mathbf{F}^{(m)}(\tau)=\lim_{\xi\rightarrow\tau} \sum_{\substack{j,l\geqslant 0\\j+l=m}}\frac{m!}{j!l!}\mathscr{M}^{(j)}(\xi)\mathbf{f}^{(l)}(\tau),\quad 0\leqslant m\leqslant\mathfrak{m}_\tau,\label{reUu1}\\
  &\mathbf{G}^{(m)}(\tau)=\lim_{\xi\rightarrow\tau} \sum_{\substack{j,l\geqslant 0\\j+l=m}}\frac{m!}{j!l!}
  \left[\mathscr{M}^{-\T}\right]^{(j)}(\xi)\mathbf{g}^{(l)}(\tau),\quad 0\leqslant m\leqslant \mathfrak{m}_\tau.\label{reVv1}
\end{alignat}
Conversely,
\begin{equation}
  \mathbf{f}^{\T}(\tau)=\lim_{\xi\rightarrow\tau}
 \mathbf{F}^{\T}(\tau) \mathscr{M}^{-\T}(\xi). \label{reUu2}
\end{equation}
 Both $ \mathscr{M}(\xi)$ and its inverse admit the representations
\begin{alignat}{1}\label{def:M4}
  &\mathscr{M}(\xi)=\mathbf{I}+\sum_{\tau\in\Omega}\sum_{\substack{j,l,m\geqslant0\\ j+l+m=\mathfrak{m}_\tau}}\frac{\mathbf{F}^{(j)}(\tau)\left[\mathbf{g}^{(l)}(\tau)\right]^\T}{j!l!(\xi-\tau)^{m+1}},\quad\xi\in\Omega,\\
  &\mathscr{M}^{-\T}(\xi)=\mathbf{I}-\sum_{\tau\in\Omega}\sum_{\substack{j,l,m\geqslant0\\ j+l+m=\mathfrak{m}_\tau}}\frac{\mathbf{G}^{(j)}(\tau)\left[\mathbf{f}^{(l)}(\tau)\right]^\T}{j!l!(\xi-\tau)^{m+1}},\quad\xi\in\Omega.\label{def:Mi5}
\end{alignat}
Inserting  \eqref{def:Mi5} into  \eqref{reVv1} and \eqref{reUu2} gives
\begin{equation}\label{regUM}
 \mathbf{f}^{\T}(\tau)=\mathbf{F}^{\T}(\tau)-(\mathscr{R}\mathbf{f}^\T)(\tau),\quad    \mathbf{G}^{(m)}(\tau)=\mathbf{g}^{(m)}(\tau)+(\mathscr{K}_m^\T\mathbf{G})(\tau), \quad 0\leqslant m\leqslant\mathfrak{m}_\tau.
\end{equation}
Let  $\mathscr{R}$  be the operator defined in   \eqref{def:scrR} with $\mathbf{F}(\tau)$ and $\mathbf{G}^{(m)}(\tau)$ supplied by  \eqref{regUM}. Repeating the argument of Theorem \ref{TH:GRS},  we obtain
 \begin{equation}\label{KRre4}
  \begin{aligned}
        &(\mathscr{R}\circ\mathscr{K})(\mathbf{h}^\T)(\xi)=    \sum_{\eta\in\Omega}\frac{1}{\mathfrak{m}_\eta!}\partial_\eta^{\mathfrak{m}_\eta}\left[\frac{(\mathscr{R}\mathbf{f}^\T)(\xi)\mathbf{g}(\eta)+\mathbf{F}^\T(\xi)(\mathscr{K}^\T\mathbf{G})(\eta)}{\xi-\eta}\mathbf{h}^\T(\eta)\right]\\
                =&\sum_{\eta\in\Omega}\frac{1}{\mathfrak{m}_\eta!}\partial_\eta^{\mathfrak{m}_\eta}\left[\frac{[\mathbf{F}^\T(\xi)-\mathbf{f}^\T(\xi)]\mathbf{g}(\eta)+\mathbf{F}^\T(\xi)[\mathbf{G}(\eta)-\mathbf{g}(\eta)]}{\xi-\eta}\mathbf{h}^\T(\eta)\right]\\
        =&\mathscr{R}(\mathbf{h}^\T)(\xi)-\mathscr{K}(\mathbf{h}^\T)(\xi),
  \end{aligned}
  \end{equation}
so that the operator $\mathbbm{1}-\mathscr{K}$  is invertible and formulae \eqref{def:scrR}, \eqref{reUu1} and \eqref{reVv1}  describe its resolvent kernel.

The equivalence of the invertibility of  $\mathbbm{1}-\mathscr{K}$
and the unique solvability of RH problem \ref{RH2} is thereby established.
\end{proof}

When $\mathfrak{m}_\tau=0$, a careful inspection of the operator \eqref{def:gdio} shows that it coincides exactly with the discrete integrable operator introduced by Borodin in \cite{Borodin2003}.

When $\mathfrak{m}_\tau=1$,  we fix the vectors
\begin{alignat}{2}
   & \mathbf{f}_0(\tau)=\mathbf{f}(\tau),\quad  &&\mathbf{f}_1(\tau)=\lim_{\xi\rightarrow\tau}\partial_\xi\mathbf{f}(\xi),\\
   & \mathbf{g}_0(\tau)=\mathbf{g}(\tau),\quad  &&\mathbf{g}_1(\tau)=\lim_{\xi\rightarrow\tau}\partial_\xi\mathbf{g}(\xi),\\
    & \mathbf{h}_0(\tau)=\mathbf{h}(\tau),\quad  &&\mathbf{h}_1(\tau)=\lim_{\xi\rightarrow\tau}\partial_\xi\mathbf{h}(\xi).
  \end{alignat}
 With these identifications the operators $\mathscr{K}$ and $\mathscr{K}_1$ given in \eqref{def:gdio} and \eqref{eq:Kjdef} reduce precisely to the matrix discrete integrable operator $\mathcal{K}$ introduced in Section~\ref{sec:sdi}.

The central novelty of the present work is to allow $\mathfrak{m}_\tau$ to be an arbitrary natural number.  This extension enlarges the family of discrete integrable operators and leads us to introduce the notion of  \emph{generalized discrete integrable operator} (See Theorem \ref{TH:GRS} and Theorem \ref{TH:GRHiif}).  The resulting class provides a broader and more flexible toolkit for integrable systems theory.

\section{ Connection to the higher-order pole solutions of integrable hierarchy }\label{sec:ihho}
\subsection{Integrable hierarhcy}
RH problems (discrete, continuous, or hybrid) are a defining feature of the inverse scattering transform for integrable PDEs. Using this structure, we can combine integrable operators with RH methods to extract integrable hierarchies of nonlinear evolution equations. The subsequent text explains how a broad class of generalized discrete integrable operators generates a complete integrable hierarchy through a single RH problem.

  Fix a positive integer $N$ and place $2N$ distinct points
   \begin{equation}
     \Omega_+=\{\xi_1,\ldots,\xi_N\}\subset\mathbb{C}^+, \quad \Omega_-=\{\bar\xi_1,\ldots,\bar\xi_N\}\subset\mathbb{C}^-,\quad \Omega=\Omega_+\cup\Omega_-.
   \end{equation}
Assign multiplicities $\mathfrak{m}_{\xi_j}=\mathfrak{m}_{\bar\xi_j}=\mathfrak{m}_j$ for $1\leqslant j\leqslant N$.
   Introduce the infinite set of time-flow parameters  $\ut=(t_1,t_2,\ldots,)$ and  the phase \begin{equation}
     \theta(\xi;\ut)=\sum_{k=1}^\infty t_k\xi^k.
   \end{equation}
Let $\mathbf{b}(\xi)\in\mathbb{C}^n$  be analytic in a neighbourhood of $\Omega$ and define
\begin{subequations}\label{def:fg1}
    \begin{equation}
      \mathbf{f}(\xi;\ut)=\begin{pmatrix}
                      \chi_{\Omega_-}(\xi) \\
                      \e^{2\ii\theta(\xi;\ut)}\mathbf{b}(\xi)\chi_{\Omega_+}(\xi)
                    \end{pmatrix},\quad \mathbf{g}(\xi;\ut)=\begin{pmatrix}
                      \chi_{\Omega_+}(\xi) \\
                    -\e^{-2\ii\theta(\xi;\ut)}\overline{\mathbf{b}(\bar\xi)}\chi_{\Omega_-}(\xi)
                    \end{pmatrix},
                    \end{equation}
 together with their higher $\xi$-derivatives
      \begin{equation}
       \partial_\xi^{j}\mathbf{f}(\xi;\ut)=\begin{pmatrix}
                      0\\
                      \partial_\xi^{j}[\e^{2\ii\theta(\xi;\ut)}\mathbf{b}(\xi)]\chi_{\Omega_+}(\xi)
                    \end{pmatrix},\quad \partial_\xi^{j}\mathbf{g}(\xi;\ut)=\begin{pmatrix}
                      0 \\
                    -\partial_\xi^{j}[\e^{-2\ii\theta(\xi;\ut)}\overline{\mathbf{b}(\bar\xi)}]\chi_{\Omega_-}(\xi)
                    \end{pmatrix}.
              \end{equation}
\end{subequations}
for $j\geqslant 1$, where $\chi$ denotes the indicator function.
\begin{definition}[Time-dependent generalized discrete integrable operator]\label{def:tddio}
  A  operator $\mathscr{K}(\ut)$ is called to be \emph{time-dependent generalized discrete integrable} if it is integrable in the sense of \eqref{def:gdio} and the associated vectors $\mathbf{f}$ and $\mathbf{g}$  depend on the time-flow parameters  $\ut=(t_1,t_2,\ldots,)$ specified in  \eqref{def:fg1}.
\end{definition}
\begin{definition}\label{def:pqrs}
Assume $\mathbbm{1}-\mathscr{K}(\ut)$ is invertible and introduce the $(n+1)\times (n+1)$ matrix
  \begin{equation}\label{eq:pqrs}
    -2\ii \sum_{\xi\in\Omega}\frac{1}{\mathfrak{m}_\xi!}\partial_\xi^{\mathfrak{m}_\xi}
    \left[\left[(\mathbbm{1}-\mathscr{K}(\ut))^{-1}\mathbf{f}\right](\xi;\ut)\mathbf{g}^\T(\xi;\ut)\right]:=\begin{pmatrix}
      p(\ut)&\mathbf{r}(\ut)\\
      \mathbf{q}(\ut)&\mathbf{S}(\ut)
    \end{pmatrix},
  \end{equation}
where  $\mathbf{q}$ is the lower-left $n\times 1$ block and $\mathbf{r}$ the upper-right $1\times n$ block.
\end{definition}

By Theorem \ref{TH:GRHiif}, the invertibility of  $\mathbbm{1}-\mathscr{K}(\ut)$  is equivalent to the unique solvability of the RH problem \ref{RH2}. For the latter, its residue condition takes the following form for  all   $j$ and  $ 0\leqslant \mathfrak{n}_j\leqslant\mathfrak{m}_{j}$,
    \begin{alignat}{1}
      &\underset{\xi=\xi_j}{\res}(\xi-\xi_j)^{\mathfrak{m}_j-\mathfrak{n}_j}\mathscr{M}_{\mathrm{L}}(\xi;\ut)=\lim_{\xi\rightarrow\xi_j}\frac{\partial_\xi^{\mathfrak{n}_j}[\e^{2\ii\theta(\xi;\ut)}\mathscr{M}_{\mathrm{R}}(\xi;\ut)\mathbf{b}(\xi)]}{\mathfrak{n}_j!},\\
      &\underset{\xi=\bar\xi_j}{\res}(\xi-\bar\xi_j)^{\mathfrak{m}_j-\mathfrak{n}_j}\mathscr{M}_{\mathrm{R}}(\xi;\ut)=-\lim_{\xi\rightarrow\bar\xi_j}\frac{\partial_\xi^{\mathfrak{n}_j}[\e^{-2\ii\theta(\xi;\ut)}\mathscr{M}_{\mathrm{L}}(\xi;\ut)\mathbf{b}^\dag(\bar\xi)]}{\mathfrak{n}_j!}.
            \end{alignat}
      \begin{theorem}\label{Th:pqrs}
        The operator  $\mathbbm{1}-\mathscr{K}(\ut)$ is invertible. The blocks $p(\ut)$, $\mathbf{q}(\ut)$,  $\mathbf{r}(\ut)$, $\mathbf{S}(\ut)$  given by Definition \ref{def:pqrs} satisfy
        \begin{equation}
        \mathbf{r}(\ut)=\mathbf{q}^\dag(\ut),\quad
  \partial_{t_1}p(\ut)=-|\mathbf{q}(\ut)|^2,\quad \partial_{t_1}\mathbf{S}(\ut)=\mathbf{q}(\ut)\mathbf{q}^\dag(\ut).
 \end{equation}
 Moreover, $\mathbf{ q}(\ut)$ obeys the $n$-component focusing NLS hierarchy,  $p(\ut)$ the KP hierarchy, and $\mathbf{S}(\ut)$
  the non-commutative KP hierarchy.
      \end{theorem}
      \begin{proof}
       The poles in RH problem \ref{RH2} are traded for jumps via an equivalence transformation that preserves the asymptotics at infinity.
Enclose  $\Omega_+$ in  a  simple closed contour $\Gamma_+\subset\mathbb{C}^+$ oriented counter-clockwise, set $\Gamma_-=\{\xi\in\mathbb{C}|\bar\xi\in\Gamma_+\}$  oriented clockwise, and let
$\Gamma=\Gamma_+\cup\Gamma_-$.

  Define
   \begin{equation}\label{def:M}
     \mathbf{M}(\xi;\ut)=\begin{cases}
       \mathscr{M}(\xi;\ut)\begin{pmatrix}
                             1 & \mathbf{0} \\
                             -\sum_{k=1}^N\frac{\e^{2\ii\theta(\xi;\ut)}\mathbf{b}(\xi)}{(\xi-\xi_k)^{m_k+1}} & \mathbf{I}
                           \end{pmatrix},\quad &\xi \text{ inside}\ \Gamma_+,\\
        \mathscr{M}(\xi;\ut)\begin{pmatrix}
                             1 & \sum_{k=1}^N\frac{\e^{-2\ii\theta(\xi;\ut)}\mathbf{b}^\dag(\bar\xi)}{(\xi-\bar\xi_k)^{m_k+1}} \\
                             \mathbf{0} & \mathbf{I}
                           \end{pmatrix},\quad &\xi  \text{  inside}\ \Gamma_-,\\
                \mathscr{M}(\xi;\ut), &\text{otherwise}.
     \end{cases}
   \end{equation}
 The function $\mathbf{M}(\xi;\ut)$ is analytic on $\mathbb{C}\backslash\Gamma$ and satisfies the jump
    \begin{equation}
      \mathbf{M}_+(\xi;\ut)= \mathbf{M}_-(\xi;\ut)\mathbf{J}(\xi;\ut),\quad \xi\in\Gamma,
    \end{equation}
    with \begin{equation}\label{eq:Mjump}
      \mathbf{J}(\xi;\ut)=\begin{pmatrix}
                             1 & -\sum_{k=1}^N\frac{\e^{-2\ii\theta(\xi;\ut)}\mathbf{b}^\dag(\bar\xi)}{(\xi-\bar\xi_k)^{m_k+1}}\chi_{\Gamma_-}(\xi) \\
                             -\sum_{k=1}^N\frac{\e^{2\ii\theta(\xi;\ut)}\mathbf{b}(\xi)}{(\xi-\xi_k)^{m_k+1}}\chi_{\Gamma_+}(\xi) & \mathbf{I}
                           \end{pmatrix}.
    \end{equation}
   Schwarz symmetry of $\mathbf{J}(\xi;\ut)$ and  Zhou’s vanishing lemma \cite{Zhou1989} guarantee  existence of the solution.  Therefore, the operator  $\mathbbm{1}-\mathscr{K}(\ut)$ is invertible.

    Note that   $\det[ \mathbf{J}(\xi;\ut)]\equiv 1$, so $\det[\mathbf{M}(\xi;\ut)]$ is an entire function of $\xi$. In view of its asymptotics at infinity, Liouville’s theorem gives $\det[\mathbf{M}(\xi;\ut)]\equiv 1$, ensuring uniqueness of the solution.  Schwarz symmetry also implies
\begin{equation}\label{eq:Msym}
  \mathbf{M}^\dag(\bar\xi;\ut)=\mathbf{M}^{-1}(\xi;\ut).
\end{equation}
 Expanding  $\mathbf{M}(\xi;\ut)$ at infinity,
 \begin{equation}\label{eq:Mle}
   \mathbf{M}(\xi;\ut)=\sum_{j=0}^\infty \mathbf{M}_j(\ut)\xi^{-j}, \quad \mathbf{M}_0=\mathbf{I},
 \end{equation}
  the Schwarz symmetry \eqref{eq:Msym} yields the anti-Hermitian constraint
 \begin{equation}\label{eq:M1sym}
\mathbf{M}_1^\dag(\ut)=- \mathbf{M}_1(\ut).
 \end{equation}
 
From \eqref{def:M3} and Definition \ref{def:pqrs}, we have
    \begin{equation}\label{recon1}
       \begin{pmatrix}
      p(\ut)&\mathbf{r}(\ut)\\
      \mathbf{q}(\ut)&\mathbf{S}(\ut)
    \end{pmatrix}=-2\ii \lim_{\xi\rightarrow \infty}\xi\left[\mathbf{M}(\xi;\ut)-\mathbf{I}\right]=-2\ii\mathbf{M}_1(\ut),
    \end{equation}
    and \eqref{eq:M1sym} immediately gives  $\mathbf{r}(\ut)=\mathbf{q}^\dag(\ut)$.

  Introduce
\begin{equation}\label{eq:Udef}
       \mathbf{U}_k(\xi;\ut)=[\partial_{t_k}\mathbf{M}(\xi;\ut)]\mathbf{M}^{-1}(\xi;\ut)-\ii\xi^{k}\mathbf{M}(\xi;\ut)\sigma\mathbf{M}^{-1}(\xi;\ut),\quad k\geqslant 1,
\end{equation}
with $\sigma=\operatorname{diag}(1,-\mathbf{I}_n)$.
A straightforward calculation shows that each   $\mathbf{U}_k(\xi;\ut)$ is holomorphic across $\Gamma$ and hence entire in $\xi$. Inspecting its  large-$\xi$ behaviour, we further conclude that
 $\mathbf{U}_k(\xi;\ut)$ is a matrix  polynomial in $\xi$ of degree at most $k$, so we may write
 \begin{equation}\label{eq:Ule}
    \mathbf{U}_k(\xi;\ut)=\sum_{j=0}^k \mathbf{u}_j(\ut)\xi^{k-j}.
 \end{equation}
 Inserting  this expansion together with   \eqref{eq:Mle} into  \eqref{eq:Udef} yields
  \begin{equation}
  \begin{aligned}
      &\sum_{j=1}^\infty \partial_{t_k}\mathbf{M}_j(\ut)\xi^{-j}-\ii\sum_{j=0}^\infty \mathbf{M}_j(\ut)\xi^{k-j}\sigma\\
      = &\sum_{j=0}^k\sum_{l=0}^j\mathbf{u}_l(\ut)\mathbf{M}_{j-l}(\ut)\xi^{k-j}
      +\sum_{j=k+1}^\infty\sum_{l=0}^k\mathbf{u}_l(\ut)\mathbf{M}_{j-l}(\ut)\xi^{k-j},\quad k\geqslant 1.
  \end{aligned}
 \end{equation}
Matching powers of $\xi$  gives
 \begin{subequations}
   \begin{alignat}{2}
     O(\xi^{k-j}):\,&-\ii\mathbf{M}_j\sigma=\sum_{l=0}^j\mathbf{u}_l\mathbf{M}_{j-l}, &&0\leqslant j
     \leqslant k,\\
     O(\xi^{-j}):\,& \partial_{t_1}\mathbf{M}_j-\ii\mathbf{M}_{j+1}\sigma=
     \mathbf{u}_0\mathbf{M}_{j+1}+\mathbf{u}_1\mathbf{M}_{j},\quad && j\geqslant 1.
   \end{alignat}
 \end{subequations}
 In particular,
 \begin{equation}
     \mathbf{u}_0=-\ii\sigma, \quad \mathbf{u}_1=\ii[\sigma,\mathbf{M}_1]=2\ii\sigma\mathbf{M}_1^{(\mathbf{O})}=\begin{pmatrix}
     0&-\mathbf{q}^\dag(\ut)\\\mathbf{q}(\ut)&\mathbf{0}
   \end{pmatrix}.
   \end{equation}
   For $l\geqslant 2$, we obtain the recursive relations
    \begin{equation}
    \begin{aligned}
    \mathbf{u}_l&=\ii[\sigma,\mathbf{M}_l]-\sum_{j=1}^{l-1}\mathbf{u}_j\mathbf{M}_{l-j}
   =-\partial_{t_1}\mathbf{M}_{l-1}-\sum_{j=2}^{l-1}\mathbf{u}_j\mathbf{M}_{l-j}\\
   &=-\partial_{t_1}\mathbf{M}_{l-1}^{(\mathbf{O})}
   -\mathbf{u}_1\mathbf{M}_{l-1}^{(\mathbf{O})}-\sum_{j=2}^{l-1}\mathbf{u}_j\mathbf{M}_{l-j},
   \end{aligned}
   \end{equation}
   together with
   \begin{equation}\label{eq:Mpr}
     \partial_{t_1}\mathbf{M}_{l-1}^{(\mathbf{D})}=\mathbf{u}_1\mathbf{M}_{l-1}^{(\mathbf{O})},\quad
   \mathbf{M}_{l-1}^{(\mathbf{O})}=-\frac{\ii}{2}\sigma\sum_{j=1}^{l-1}\mathbf{u}_j\mathbf{M}_{l-1-j},
   \end{equation}
where the superscripts  $(\mathbf{D})$ and $(\mathbf{O})$  denote the block-diagonal and block-off-diagonal parts of an  $(n+1)\times (n+1)$ matrix  partitioned as
\begin{equation}
 \left(
\begin{array}{c|ccc}
               a_{11} & a_{12}&\cdots&a_{1,n+1}\\\hline
               a_{21} & a_{22}&\cdots&a_{2,n+1}\\
               \vdots&\vdots&&\vdots\\
               a_{n+1,1} & a_{n+1,2}&\cdots&a_{n+1,n+1}
             \end{array}\right).
\end{equation}
Carrying out the recursion one finds
\begin{equation}
  \begin{aligned}
      \mathbf{u}_2&=-\partial_{t_1}\mathbf{M}_1^{(\mathbf{O})}-\mathbf{u}_1\mathbf{M}_1^{(\mathbf{O})}=\frac{\ii}{2}\sigma(\partial_{t_1}\mathbf{u}_1-\mathbf{u}_1^2),\\
    \mathbf{u}_3&=-\partial_{t_1}\mathbf{M}_2^{(\mathbf{O})}-\mathbf{u}_1\mathbf{M}_2^{(\mathbf{O})}-\mathbf{u}_2\mathbf{M}_1=\frac{1}{4}\left([\partial_{t_1}\mathbf{u}_1,\mathbf{u}_1]-\partial_{t_1}^2\mathbf{u}_1+2\mathbf{u}_1^3\right).
  \end{aligned}
\end{equation}
Finally, \eqref{eq:M1sym} and \eqref{eq:Mpr} imply $\partial_{t_1}p(\ut)=-|\mathbf{q}(\ut)|^2$ and $\partial_{t_1}\mathbf{S}(\ut)=\mathbf{q}(\ut)\mathbf{q}^\dag(\ut)$.

Introduce the wave function
\begin{equation}
  \Psi(\xi;\ut)=\mathbf{M}(\xi;\ut)\e^{-\ii\theta(\xi;\ut)\sigma}.
\end{equation}  Equation $\eqref{eq:Udef}$ shows that $\Psi(\xi;\ut)$ satisfies the linear  compatible system
\begin{equation}
  \partial_{t_k}\Psi(\xi;\ut)=\mathbf{U}_k(\xi;\ut)\Psi(\xi;\ut),\quad k\geqslant 1,
\end{equation}
whose  integrability is enforced by the zero curvature equations
\begin{equation}\label{eq:zce}
  \partial_{t_j}\mathbf{U}_k-\partial_{t_k}\mathbf{U}_j+\left[\mathbf{U}_k,\mathbf{U}_j\right]=\mathbf{0},\quad j\neq k.
   \end{equation}

Specializing the flow parameters to $\ut=(x,2y,0,0,\ldots)$ and $\ut=(x,0,4t,0,\ldots)$,   respectively, from the zero curvature equation \eqref{eq:zce}  we recover the $n$-component  focusing NLS equation
\begin{equation}\label{vfnls}
  -\mathrm{i} \partial_y\mathbf{q}+ \partial_x^2\mathbf{q}+2 | \mathbf{q}|^2 \mathbf{q}= \mathbf{0},
\end{equation} and the $n$-component complex  mKdV equation
\begin{equation}\label{vmkdv}
 \partial_t\mathbf{q}+  \partial_x^3\mathbf{q}+3 | \mathbf{q}|^2 \partial_x\mathbf{q}+3  \mathbf{q}\mathbf{q}^\dag\partial_x\mathbf{q}= \mathbf{0}.
\end{equation}
 A direct calculation shows that the scalar potentials
\begin{equation}
   p_\pm(x,y,t)=|\mathbf{ q}(x,2\sqrt{\pm1}y,4t,0,\ldots)|^2=-\partial_xp(x,2\sqrt{\pm1}y,4t,0,\ldots)
\end{equation}  satisfy the KP-I and KP-II equations
\begin{equation}\label{eq:KP}
  4\partial_x\partial_t p_\pm+\partial_x^4p_\pm+6\partial_x^2(p_\pm^2)\mp3\partial_y^2p_\pm=0,
\end{equation}\label{eq:KP-II}
 while the matrix  potential
 \begin{equation}
   \mathbf{Q}(x,y,t)=\mathbf{ q}(x,2\ii y,4t,0,\ldots)\mathbf{ q}^\dag(x,2\ii y,4t,0,\ldots)=\partial_x\mathbf{S}(x,2\ii y,4t,0,\ldots)
 \end{equation} solves the non-commutative KP-II equation
\begin{equation}\label{eq:NKP-II}
   4\partial_x\partial_t\mathbf{Q}+\partial_x^4\mathbf{Q}+6\partial_x^2(\mathbf{Q}^2)+3\partial_y^2\mathbf{Q}+6\partial_x[\mathbf{Q},\partial_y\mathbf{S}]=\mathbf{0}.
\end{equation}
The proof is now complete.
  \end{proof}
  \subsection{Higher-order pole solution}
As previously discussed, the generalized integrable operator extends the RH formalism  beyond pure soliton solutions, enabling the incorporation of higher-order pole singularities. In this work, we illustrate this paradigm by constructing higher-order pole solutions of the $2$-component focusing  NLS hierarchy

 By  Definition \ref{def:pqrs} and Theorem \ref{Th:pqrs}, the solution of the $2$-component focusing  NLS hierarchy can be expressed as
\begin{equation}
  \mathbf{q}(\ut)=-2\ii \sum_{\xi\in\Omega_+}\frac{1}{\mathfrak{m}_\xi!}\partial_\xi^{\mathfrak{m}_\xi}
    \left[\mathbf{B}(\xi;\ut)\right],
\end{equation}
where
\begin{equation}
\mathbf{B}(\xi;\ut)=(\mathbbm{1}-\mathscr{K}(\ut))^{-1}[\e^{2\ii\theta(\xi;\ut)}\mathbf{b}(\xi)\chi_{\Omega_+}(\xi)].
\end{equation}
That is, for each $\xi\in\Omega$, the functions $\{\mathbf{B}(\xi;\ut),\ldots,\partial_\xi^{\mathfrak{m}_\xi}\mathbf{B}(\xi;\ut)\}$ satisfies the following linear system:
\begin{equation}\label{eq:sym}
  \begin{cases}
    \mathbf{B}(\xi;\ut)-\mathscr{K}(\ut) \mathbf{B}(\xi;\ut)=\e^{2\ii\theta(\xi;\ut)}\mathbf{b}(\xi)\chi_{\Omega_+}(\xi),\\
    \qquad\qquad\vdots\\
    \partial_\xi^{\mathfrak{m}_\xi}\mathbf{B}(\xi;\ut)-\partial_\xi^{\mathfrak{m}_\xi}[\mathscr{K}(\ut) \mathbf{B}(\xi;\ut)]=\partial_\xi^{\mathfrak{m}_\xi}\left[\e^{2\ii\theta(\xi;\ut)}\mathbf{b}(\xi)\right]\chi_{\Omega_+}(\xi).
  \end{cases}
\end{equation}
Substituting \eqref{def:fg1} into \eqref{eq:sym},
we find that $\{\mathbf{B}(\xi;\ut),\ldots,\partial_\xi^{\mathfrak{m}_\xi}\mathbf{B}(\xi;\ut):\xi\in\Omega_+\}$ satisfies the following linear systems:
\begin{equation}
  \begin{cases}
   \displaystyle \mathbf{B}(\xi;\ut)+\sum_{\eta\in\Omega_+}\sum_{\zeta\in\Omega_-}\partial_\zeta^{\mathfrak{m}_\zeta}\partial_\eta^{\mathfrak{m}_\eta}
   \frac{\e^{2\ii[\theta(\xi;\ut)-\theta(\zeta;\ut)]}\mathbf{b}^\T(\xi)\overline{\mathbf{b}(\bar\zeta)}\mathbf{B}(\eta;\ut)}{(\xi-\zeta)(\zeta-\eta)}=\e^{2\ii\theta(\xi;\ut)}\mathbf{b}(\xi),\quad \xi\in\Omega_+,\\
    \qquad\qquad\vdots\\
    \displaystyle\partial_\xi^{\mathfrak{m}_\xi}\mathbf{B}(\xi;\ut)+\sum_{\eta\in\Omega_+}\sum_{\zeta\in\Omega_-}\partial_\xi^{\mathfrak{m}_\xi}\partial_\zeta^{\mathfrak{m}_\zeta}\partial_\eta^{\mathfrak{m}_\eta}
   \frac{\e^{2\ii[\theta(\xi;\ut)-\theta(\zeta;\ut)]}\mathbf{b}^\T(\xi)\overline{\mathbf{b}(\bar\zeta)}\mathbf{B}(\eta;\ut)}{(\xi-\zeta)(\zeta-\eta)}=\partial_\xi^{\mathfrak{m}_\xi}\left[\e^{2\ii\theta(\xi;\ut)}\mathbf{b}(\xi)\right].
  \end{cases}
\end{equation}

As a concrete example,  let  $N=1$, $\xi_1=\ii$,  $\mathfrak{m}_1=1$, and $\mathbf{b}(\xi)=(\xi+\ii,1)^\T$. Then $\Omega_+=\{\ii\}$, $\Omega_-=\{-\ii\}$. When $\ut=(x,2y,0,0,\ldots)$, the second-order pole solution of the $2$-component focusing  NLS equation \eqref{vfnls} is given by
\begin{equation}
  \mathbf{q}(\ut)= \begin{pmatrix}\displaystyle
   \frac{32 \ii \e^{2 x-4 \ii y} \left[16 \e^{4 x} (16 \ii y+4 x-1)+96 \ii y-24 x-19\right]}{32 \e^{4 x} \left[640 y^2+8 \e^{4 x}+8 x (5 x+3)+9\right]+29}\\[3ex]\displaystyle
 \frac{64 \e^{2 x-4 \ii y} \left[16 \e^{4 x} (x+4 \ii y)+4 \ii y-x-2\right]}{32 \e^{4 x} \left[640 y^2+8 \e^{4 x}+8 x (5 x+3)+9\right]+29}
  \end{pmatrix}.
\end{equation}
When $\ut=(x,0,4t,0,\ldots)$, the second-order pole solution of the $2$-component complex mKdV equation \eqref{vmkdv}    becomes
\begin{equation}
  \mathbf{q}(\ut)=\begin{pmatrix}
                    \displaystyle \frac{32 \ii \e^{8 t+2 x} \left[\e^{16 t} (-288 t+24 x+19)+16 \e^{4 x} (48 t-4 x+1)\right]}{32 \e^{4 (4 t+x)} \left[5760 t^2-96 t (10 x+3)+8 x (5 x+3)+9\right]+29 \e^{32 t}+256 \e^{8 x}}\\[3ex]
                    \displaystyle  \frac{1024 \e^{8 t+6 x} (x-12 t)-64 \e^{24 t+2 x} (-12 t+x+2)}{32 \e^{4 (4 t+x)} \left[5760 t^2-96 t (10 x+3)+8 x (5 x+3)+9\right]+29 \e^{32 t}+256 \e^{8 x}}
                  \end{pmatrix}.
\end{equation}
Finally, when $\ut=(x,2y,4t,0,\ldots)$, the second-order pole solution of the KP-I equation \eqref{eq:KP} is given by $p_+(x,y,t)=\frac{p_1(x,y,t)}{p_2(x,y,t)}$, where
\begin{equation}
\begin{aligned}
  p_1(x,y,t)=&1024 \e^{-4 (4 t+x)} \left\{256 \e^{8 x} \left[2880 t^2+96 t (1-5 x)+4 x (5 x-2)+320 y^2+1\right]\right.\\
  &+29 \e^{32 t} \left[2880 t^2-96 t (5 x+4)+4 x (5 x+8)+320 y^2+13\right]\\
  &\left.-32 \e^{4 (4 t+x)} \left[14400 t^2-240 t (10 x+3)+20 x (5 x+3)-1600 y^2-19\right]\right\},\\
  p_2(x,y,t)=&\left[32 \left(5760 t^2-96 t (10 x+3)+8 x (5 x+3)+640 y^2+9\right)\right.\\
  &\left.-227 \sinh (16 t-4 x)+285 \cosh (16 t-4 x)\right]^2.
\end{aligned}
\end{equation}
\section{Concluding remarks}
In this work, we successfully constructed discrete integrable operators in both matrix and differential forms. We  established their intrinsic integrable structures and derived the associated RH problems that capture their integrability. A key result is that these generalized discrete integrable operators are not just abstract concepts. They are  deeply connected to classical multi-component integrable hierarchies (e.g., the NLS and the KP hierarchies) and can be explicitly represented as higher-order poles solutions. Furthermore, these generalized discrete integrable operators naturally expand the scope of matrix integrable systems.

Of particular interest is the potential use of these operators. The multi-point, multi-time, multi-position distribution structure of the KPZ universality class fixed points is inherently linked to integrable systems theory, with the matrix KP-II equation being the fundamental multi-time dynamical equation revealed through the analysis of cubic integrable operators \cite{BLS2022,BPS2023,QR2022}. We ask whether they may directly link to periodic KPZ fixed points, discrete  kernels, symplectic ensembles, and skew-orthogonal polynomials. Exploring these connections goes beyond academic interest: it could provide deeper insights into how integrable structures support phenomena ranging from the statistical mechanics of growing interfaces to the spectral statistics of complex quantum systems. This is an interesting direction for future research.

\section*{Acknowledgment}
	This work was sponsored by National Natural Science Foundation of China (Grant No. 12571268) and  Program for Science \& Technology Innovation Talents in Universities of Henan Province (Grant No. 26HASTIT049).	
	
	\section*{Data availability}
	All data generated or analyzed during this study are including in this published article.	
	
	\section*{Declarations}	
	
	\section*{Conflict of Interest}		
	The authors declare that they have no conflict of interest.


\begin{thebibliography}{99}
\bibitem{IIKS1990}Its, A. R., Izergin, A. G., Korepin, V. E.,  Slavnov, N. A. (1990). Differential equations for quantum correlation functions. International Journal of Modern Physics B, 4(05), 1003-1037.
    \bibitem{DP1993} Deift, P.,  Zhou, X. (1993). A steepest descent method for oscillatory Riemann–Hilbert problems. Asymptotics for the MKdV equation. Annals of Mathematics, 137(2), 295-368.
\bibitem{IIKS1993}Its, A. R., Izergin, A. G., Korepin, V. E.,  Slavnov, N. A. (1993). The Quantum Correlation Function as the $\tau$  Function of Classical Differential Equations. In Important developments in soliton theory (pp. 407-417). Berlin, Heidelberg: Springer Berlin Heidelberg.
    \bibitem{HI2002}Harnad, J.,   Its, A. R. (2002). Integrable Fredholm operators and dual isomonodromic deformations. Communications in Mathematical Physics, 226(3), 497-530.        
         \bibitem{BGO2024}Bertola, M., Grava, T.,  Orsatti, G. (2024). Integrable operators, $\overline {\partial} $-problems, KP and NLS hierarchy. Nonlinearity, 37(8), 085008. 
\bibitem{Borodin2000}Borodin, A. (2000). Riemann--Hilbert problem and the discrete Bessel kernel. International Mathematics Research Notices, 2000(9), 467-494.

\bibitem{BOO2000}Borodin, A., Okounkov, A.,  Olshanski, G. (2000). Asymptotics of Plancherel measures for symmetric groups. Journal of the American Mathematical Society, 13(3), 481-515.

\bibitem{BO2000}Borodin, A., Olshanski, G. (2000). Distributions on Partitions, Point Processes, and the Hypergeometric Kernel. Communications in Mathematical Physics, 211(2), 335-358.
\bibitem{BD2002}Borodin, A.,  Deift, P. (2002). Fredholm determinants, Jimbo--Miwa--Ueno $\tau$‐functions, and representation theory. Communications on Pure and Applied Mathematics, 55(9), 1160-1230.
\bibitem{Borodin2003}Borodin, A.  (2003). Discrete gap probabilities and discrete Painlevé equations. Duke Mathematical Journal, 117(3), 489-542.
\bibitem{CR2023}Cafasso, M.,  Ruzza, G. (2023). Integrable equations associated with the finite-temperature deformation of the discrete Bessel point process. Journal of the London Mathematical Society, 108(1), 273-308.
\bibitem{BLS2022}Baik, J., Liu, Z., Silva, G. L. (2022). Limiting one-point distribution of periodic TASEP. In Annales de l'Institut Henri Poincare (B) Probabilites et statistiques, 58(1), 248-302.
\bibitem{BPS2023}Baik, J., Prokhorov, A.,  Silva, G. L. (2023). Differential equations for the KPZ and periodic KPZ fixed points. Communications in Mathematical Physics, 401(2), 1753-1806.
    \bibitem{QR2022}Quastel, J.,  Remenik, D. (2022). KP governs random growth off a 1-dimensional substrate. Forum of Mathematics, Pi, 10, e10.
\bibitem{Krajenbrink2020}Krajenbrink, A. (2020). From Painlevé to Zakharov--Shabat and beyond: Fredholm determinants and integro-differential hierarchies. Journal of Physics A: Mathematical and Theoretical, 54(3), 035001.
\bibitem{MQR2025}Matetski, K., Quastel, J.,  Remenik, D. (2025). Polynuclear growth and the Toda lattice. Journal of the European Mathematical Society. DOI 10.4171/JEMS/1558.
\bibitem{CCR2022}Charlier, C., Claeys, T.,  Ruzza, G. (2022). Uniform tail asymptotics for Airy kernel determinant solutions to KdV and for the narrow wedge solution to KPZ. Journal of Functional Analysis, 283(8), 109608.
    \bibitem{KL2021}Krajenbrink, A., Le Doussal, P. (2021). Inverse scattering of the Zakharov--Shabat system solves the weak noise theory of the Kardar--Parisi--Zhang equation. Physical Review Letters, 127(6), 064101.
    \bibitem{MMS2022}Mallick, K., Moriya, H.,  Sasamoto, T. (2022). Exact solution of the macroscopic fluctuation theory for the symmetric exclusion process. Physical Review Letters, 129(4), 040601.
  
    \bibitem{GP2002}Ghosh, S.,  Pandey, A. (2002). Skew-orthogonal polynomials and random-matrix ensembles. Physical Review E, 65(4), 046221.
\bibitem{BS2009}Borodin, A.,   Strahov, E. (2009). Correlation kernels for discrete symplectic and orthogonal ensembles. Communications in Mathematical Physics, 286(3), 933-977.
 \bibitem{Strahov2010a}Strahov, E. (2010). The $Z$-measures on partitions, Pfaffian point processes, and the matrix hypergeometric kernel. Advances in Mathematics, 224(1), 130-168.
 \bibitem{Strahov2010b}Strahov, E. (2010).  $Z$-measures on partitions related to the infinite Gelfand pair $(S(2\infty),H(\infty))$. Journal of Algebra, 323(2), 349-370.
    \bibitem{Forrester2010}Forrester, P. J. (2010). Log-gases and random matrices (LMS-34). Princeton university press.
\bibitem{Petrov2011}Petrov, L. (2011). Pfaffian stochastic dynamics of strict partitions. Electronic Journal of Probability, 16, 2246–2295.
\bibitem{FM2012}Forrester, P. J.,  Mays, A. (2012). Pfaffian point process for the Gaussian real generalised eigenvalue problem. Probability Theory and Related Fields, 154(1), 1-47.
\bibitem{Kargin2014}Kargin, V. (2014). On Pfaffian random point fields. Journal of Statistical Physics, 154(3), 681-704.
\bibitem{BBCS2018}Baik, J., Barraquand, G., Corwin, I.,  Suidan, T. (2018). Pfaffian Schur processes and last passage percolation in a half-quadrant. The Annals of Probability, 46(6), 3015–3089.

\bibitem{FL2020}Forrester, P.,  Li, S. H. (2020). Classical discrete symplectic ensembles on the linear and exponential lattice: skew orthogonal polynomials and correlation functions. Transactions of the American Mathematical Society, 373(1), 665-698.
\bibitem{Widom1999}Widom, H. (1999). On the relation between orthogonal, symplectic and unitary matrix ensembles. Journal of Statistical Physics, 94(3), 347-363.
\bibitem{Little2024} Little, A. (2024). A Riemann-Hilbert approach to skew-orthogonal polynomials of symplectic type. Symmetry, Integrability and Geometry: Methods and Applications, 20(076), 32.
\bibitem{IK2014}Its, A. R.,  Kozlowski, K. K. (2014). On determinants of integrable operators with shifts. International Mathematics Research Notices, 2014(24), 6826-6838.
\bibitem{IK2016}Its, A. R.,  Kozlowski, K. K. (2016). Large-$x$ analysis of an operator-valued Riemann--Hilbert problem. International Mathematics Research Notices, 2016(6), 1776-1806.

\bibitem{LSG20241}Liu, H., Shen, J.,  Geng, X. G. (2024). Multiple higher-order pole solutions in spinor Bose--Einstein condensates. Journal of Nonlinear Science, 34(3), 48.
\bibitem{LSG20242}Liu, H., Shen, J.,  Geng, X. G. (2024). Riemann–Hilbert method to the Ablowitz--Ladik equation: Higher-order cases. Studies in Applied Mathematics, 153(3), e12748.
\bibitem{Li2025}Li, S. T. (2025). On Zero-Background Solitons of the Sharp-Line Maxwell–Bloch Equations. Communications in Mathematical Physics, 406(4), 89.
\bibitem{Zhou1989}Zhou, X. (1989). The Riemann--Hilbert problem and inverse scattering. SIAM journal on mathematical analysis, 20(4), 966-986.


\end{thebibliography}
\end{document}